\documentclass{aa}  

\usepackage[dvipsnames]{xcolor}
\usepackage{graphicx}
\usepackage{txfonts}
\usepackage{hyperref}
\hypersetup{
    colorlinks=true,
    linkcolor=blue,
    urlcolor=magenta,
    citecolor=blue,    
}
\usepackage{listings}
\definecolor{dkgreen}{rgb}{0,0.6,0}
\definecolor{gray}{rgb}{0.5,0.5,0.5}
\definecolor{mauve}{rgb}{0.58,0,0.82}

\usepackage{placeins}
\newcommand{\orcidlink}[1]{\protect\href{https://orcid.org/#1}{\protect\includegraphics[width=8pt]{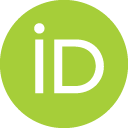}}}

\newcommand{\ThorSimName}[0]{{NEXUS\,}}

\begin{document}

   \title{Signatures of simulated spiral arms on radial actions}


   \author{P. A. Palicio \inst{1}\thanks{email to: pedro.alonso-palicio@oca.eu}\orcidlink{0000-0002-7432-8709}  \and A. Recio-Blanco \inst{1} \and T. Tepper-Garcia \inst{2,3} \and E. Poggio \inst{4} \and S. Peirani \inst{5,6,1,7} \and\\ Y. Dubois \inst{7} \and  P. J. McMillan \inst{8} \and J. Bland-Hawthorn \inst{2,3} \and K. Kraljic \inst{9}  \and M. Barbillon \inst{1}
          }

   \institute{
   \inst{1} {Universit\'e C\^ote d'Azur, Observatoire de la C\^ote d'Azur, CNRS, Laboratoire Lagrange, Bd de l'Observatoire,  CS 34229, 06304 Nice cedex 4, France}\\
    \inst{2} {Sydney Institute for Astronomy, School of Physics, A28, The University of Sydney, NSW 2006, Australia}\\
    \inst{3} {Center of Excellence for All Sky Astrophysics in Three Dimensions (ASTRO-3D), Australia}\\
    \inst{4} {INAF - Osservatorio Astrofisico di Torino, via Osservatorio 20, 10025 Pino Torinese (TO), Italy}\\
    \inst{5} {ILANCE, CNRS – University of Tokyo International Research Laboratory, Kashiwa, Chiba 277-8582, Japan}\\
    \inst{6} {Kavli IPMU (WPI), UTIAS, The University of Tokyo, Kashiwa, Chiba 277-8583, Japan}\\
    \inst{7} {Sorbonne Université, CNRS, UMR7095, Institute d' Astrophysique de Paris, 98 bis Boulevard Arago, 75014 Paris, France}\\
    \inst{8} {School of Physics \& Astronomy, University of Leicester, University Road, Leicester LE1 7RH, UK.}\\
    \inst{9} {Universit\'e de Strasbourg, CNRS, Observatoire astronomique de Strasbourg, UMR 7550, F-67000 Strasbourg, France}
    }
    \date{Received XXXX; accepted YYY}


\abstract
   {Spiral arms play a key role in the evolution of disc galaxies, defining their morphology, star formation, chemistry, and dynamics. Among their various implications, it has been observed in the Milky Way disc that the distribution of Gaia Data Release 3 (DR3) radial actions exhibits structures that might be related to the spiral arms.}
   {Our goal is to investigate the relationship between regions of low radial action identified in simulated discs and the location of the spiral arms, such as that {suggested for the Galaxy in previous studies}.}
   {For a sample of 23 simulated spiral galaxies, we modelled the axisymmetric component of their gravitational potential to compute the radial action of their stellar particles using the Stäckel fudge. The spatial distribution of the radial action was then compared to the location of the spiral arms, identified as overdensities in the stellar surface density using a kernel density estimator.}
   {Our analysis reveals a strong correlation between the radial action distribution and the spiral arms in 18 of 23 simulated galaxies. However, notable discrepancies are observed in the remaining five, since they are profoundly out-of-equilibrium systems, such as galaxies influenced by external interactions or spiral arms still in the process of winding up. Additionally, spiral arms are consistently traced across young, intermediate, and old stellar populations ($\geq3$~Gyr) in most simulations, suggesting that they are supported by stars spanning a broad age range.}
    {We have confirmed that, in general, there is a tendency for spatial correlation between spiral arms and stellar populations featuring low values of the radial action, {as discussed in the literature using Gaia DR3 data}. However, discrepancies between features in the radial action distribution and the spiral structure can be interpreted as signatures of recent disturbances, a scenario applicable to the Milky Way. Furthermore, populations at least as old as $3$~Gyr trace the spiral arms with no significant misalignment across age bins, suggesting a possible theoretical interpretation of the observations obtained with Gaia data. A linear relation between the maximum value of the radial action of the spiral arms and the vertical scale-length is found, which is also satisfied by the Milky Way.}

   \keywords{Galaxy: kinematics and dynamics --
                Galaxy: structure --
                Galaxy: disk
               }

   \authorrunning{Palicio et al.}

   \maketitle
%
\section{Introduction}
\par Spiral arms are ubiquitous structures in the Local Universe, present in about two-thirds of galaxies \citep{HuertasCompany2011, Lintott2011, Willett2013, Kelvin2014, Hart2016, Hart2017a, Masters2019}. This prevalence has facilitated the morphological classification of spiral arms into three categories \citep{Elmegreen82}: the `grand design', typically featuring two large, continuous, well-defined arms that complete one revolution around the galaxy; flocculent spiral arms, characterised by a more irregular pattern composed of numerous fragmented segments with no apparent correlation among them \citep{Goldreich1965, Binney2008}; and a third category comprising intermediate-scale arms, whose tracks exhibit bifurcations, branching, and are less extended than the `grand design' arms.  
\par Despite significant research over the last six decades, the formation mechanisms of spiral arms remain unclear, though several models have been proposed to explain these structures. These include the quasi-steady density wave theory \citep{Lindblad63, LinShu1964, BertinLin1996, Shu16}, the swing amplification model \citep{Goldreich1965, Julian1966, Toomre1981,Grand12a, Grand12b, Grand13, Dobbs2014, Michikoshi16}, transient spiral modes \citep{SellwoodCarlberg14}, the dressed mass clump model \citep{Toomre91, DOnghia13}, tidal interactions \citep{Toomre72, Tully74, Elmegreen83, Oh08, Dubinski08, Dobbs10}, and recurrent cycles of groove modes \citep{Sellwood2019} among others. 
\par As important drivers of secular evolution \citep{Kormendy04}, spiral arms not only define the morphology of galaxies but also contribute to the redistribution of angular momentum \citep{LyndenBell1972,Minchev10, Minchev11}, the heating of the disc \citep{MartinezMedina2015, Grand16}, radial migration \citep{Sellwood2002, DiMatteo13}, and the thickening of the disc \citep{Rovskar13}, to cite a few examples. Therefore, the presence of spiral arms has implications for disc kinematics and dynamics. In contrast to face-on external galaxies, where the presence of the spiral structure can be detected by direct imaging, our position within the Galactic disc complicates the detection of the Milky Way's spiral arms by this technique, requiring more indirect methods such as those based on distributions of young objects \citep{Reid14, Reid19, Xu2018, Xu2021b, Xu2021a,Pantaleoni2021, Zari21, PVPChemCart, PVPDrimmel,Ge24, RezaeiKh24}, chemistry \citep{Poggio2022, Spitoni23, Hawkins2023, Barbillon2024, Hackshaw2024, Debattista2024}, kinematics \citep{Antoja11,Hunt2015,MartinezMedina22,Denyshchenko2024}, and dynamics \citep{Trick17, Hunt2019, SellwoodTrick2019, Palicio2023,Khalil2024}. \citet{SellwoodTrick2019} performed a detailed analysis of the Gaia Data Release 2 (DR2) dynamics \citep{GaiaDR2} in the very solar neighbourhood ($\sim 200$~pc from the Sun) to test some of the spiral arm models mentioned above, concluding that the dressed mass model is favoured. \citet{MataChavez2019} studied the relation between the physical parameters of simulated spiral arms with their formation mechanisms, although they did not perform a comparison with those of the Milky Way.
\par Using the full kinematics sample from Gaia DR3 \citep{GaiaDR3}, in \citet{Palicio2023} we mapped the radial actions, $J_R$, and identified areas in the Galactic disc characterised by low values of $J_R$ compared to the whole explored region. The location and shape of most of these areas were compatible with those believed to trace spiral arms identified using young tracers such as masers \citep{Reid14}, upper main sequence stars \citep{Poggio21}, and Cepheids \citep{Lemasle22}. This correlation can be attributed to the fact that stars are believed to form on nearly circular orbits within spiral arms, resulting in low radial actions. Although satisfactory, we could not completely rule out the contribution of moving groups to the features seen in the $J_R$ maps, and some discrepancies were observed in the \textsc{IV} quadrant ($-90^{\circ}<\ell<0^{\circ}$). Our results also suggested that older populations can support the structure of spiral arms, in alignment with the findings of \citet{Lin2022} and \citet{Uppal23} concerning the distribution of red clump stars in the Local arm and the Outer arm, respectively.
\par In this work, we aim to confirm the relationship between spiral arms and regions of low $J_R$, originally reported in \citet{Palicio2023}, through numerical simulations. We also address whether older populations can trace spiral arms. This paper is organised as follows. In Section \ref{Sect_Simul}, we present the set of simulations used in this work and describe the fitting procedure for the potential and acceleration field. Results and discussions are detailed in Section \ref{Sect_Results}. Conclusions are summarised in Section \ref{Sect_Conclu}.
%

\section{Simulations}
\label{Sect_Simul}
\subsection{Selected sample and methodology}
\par We analysed a sample of 23 disc galaxy simulations at redshift zero, comprising seven from the Auriga simulations \citep{Grand17, Grand24}, ten from the Illustris-TNG collaboration \citep{Vogelsberger14, Pillepich19, Nelson2019, Nelson2019b, Pillepich24}, one from the NewHorizon collaboration \citep{Dubois_2021}, and the g2.79e12 \textsc{NIHAO-UHD} simulation \citep{Buck20}. These simulations represent Milky Way-like galaxies and were selected for the pronounced spiral arms shown in their pre-visualisation images. {The galaxies provided by the Auriga, Illustris-TNG, and NIHAO-UHD teams belong to a group or cluster of galaxies and, consequently, their evolution is also affected by interactions with their neighbours, as evidenced by the visualisation of their snapshots. We noted, however, that certain relative isolation criteria were applied to consider them as Milky Way-like candidates, as reported in their respective works. In contrast, the NewHorizon volume was chosen so that most galaxies evolve in a field environment, as does the one selected in this work}. Apart from these cosmological simulations, we have considered four controlled simulations of idealised galaxies, two of which were presented in \citet{BlandHawthorn2021} and \citet{tep22x}, respectively, and two unpublished simulations created with the {\sc Nexus} framework \citep{tep24a}. These controlled simulations include the interaction with a point mass intended as a proxy for the Sagittarius dwarf galaxy (Sgr), and differ from one another in their gas content and how it is treated: \textsc{\ThorSimName~A} \citep[cf.][]{BlandHawthorn2021} is a pure $N$-body simulation without gas, \textsc{\ThorSimName~B} and \textsc{\ThorSimName~D} \citep[cf.][]{tep22x} include an `inert' gaseous disc component with a mass equal to 20\% and 10\% the the total (stars and gas) disc mass, respectively. In these simulations, the gas is treated with a simple isothermal equation of state. Finally,  \textsc{\ThorSimName~C} adopts the same initial conditions as \textsc{\ThorSimName~B}, but it includes a prescription for galaxy formation, whereby the gas is allowed to cool and to form stars, accounting for their energetic and mass feedback. We kindly refer to the above mentioned references for more details on each the simulation suites. The nominal resolutions for the baryonic particles, as reported for each family of simulations in their respective works, are: 369~pc (\textsc{Auriga}), 150~pc (\textsc{Illustris}), 34~pc (\textsc{NewHorizon}), 36~pc (\textsc{NEXUS}), and 265~pc (\textsc{NIHAO-UHD}). The number of stellar particles contained in each simulation, as well as other relevant details, are presented in Table \ref{tab_simparam}.

\begin{table*}[ht]
	\centering
	\caption{Parameters of the simulations considered in this work. }
	\label{tab_simparam}
	\begin{tabular}{lcccccccccc}
\hline
SimID & $\rm N_{part}$ & $r_{min}$ & $a$ & $r_{max}$ & $N$ & $L$ & $h$ & $H$ & $L_{kpc}$ & $Z_{lim}$ \\
 & ($\rm \times 10^6$) & (kpc) & (kpc) & (kpc) &  &  & (kpc) & (kpc) & (kpc) & (kpc) \\
 \hline
Auriga 2 & 3.78 & 2.0 & 5.0 & 40 & 10 & 10 & 0.50 & 2.5 & 30 & 2.5 \\
Auriga 3 & 2.79 & 2.0 & 5.0 & 40 & 8 & 8 & 0.50 & 2.0 & 20 & 1.0 \\
Auriga 16 & 5.20 & 2.5 & 5.0 & 50 & 12 & 8 & 0.50 & 2.0 & 20 & 1.0 \\
Auriga 19 & 2.44 & 2.0 & 5.0 & 40 & 12 & 8 & 0.50 & 2.0 & 15 & 0.5 \\
Auriga 23 & 2.90 & 2.5 & 5.0 & 45 & 10 & 10 & 0.50 & 2.5 & 20 & 1.0 \\
Auriga 24 & 5.34 & 2.0 & 5.0 & 40 & 12 & 8 & 0.50 & 2.0 & 20 & 1.0 \\
Auriga 27 & 3.55 & 2.5 & 5.0 & 40 & 8 & 8 & 0.50 & 2.0 & 20 & 1.0 \\
Illustris 424288 & 2.15 & 1.5 & 5.0 & 40 & 16 & 12 & 0.75 & 4.0 & 20 & 3.5 \\
Illustris 456326 & 2.47 & 2.0 & 5.0 & 40 & 16 & 12 & 0.50 & 2.0 & 20 & 1.0 \\
Illustris 464163 & 1.82 & 1.0 & 5.0 & 40 & 12 & 8 & 0.75 & 4.0 & 30 & 2.5 \\
Illustris 479290 & 1.71 & 1.0 & 6.0 & 50 & 14 & 10 & 0.75 & 2.5 & 25 & 1.5 \\
Illustris 509091 & 1.00 & 2.5 & 5.0 & 40 & 14 & 10 & 0.75 & 2.5 & 25 & 2.0 \\
Illustris 532760 & 1.13 & 2.5 & 5.0 & 40 & 10 & 8 & 0.75 & 2.5 & 25 & 2.0 \\
Illustris 537941 & 0.88 & 2.5 & 4.0 & 35 & 10 & 10 & 0.50 & 2.0 & 20 & 2.0 \\
Illustris 549516 & 0.73 & 1.5 & 4.0 & 50 & 14 & 10 & 0.75 & 4.0 & 30 & 1.5 \\
Illustris 552414 & 1.00 & 2.0 & 5.0	& 40 & 10 & 10 & 0.50 & 2.0 & 20 & 1.0 \\
Illustris 555287 & 0.65 & 2.0 & 2.5 & 35 & 10 & 10 & 0.50 & 2.0 & 20 & 1.0 \\
NewHorizon & 8.98 & 0.0 & 3.0 & 40 & 14 & 14 & 0.50 & 3.0 & 25 & 1.0 \\
\ThorSimName~A & 5.00 & 2.0 & 5.0 & 40 & 10 & 10 & 0.50 & 2.0 & 30 & 1.0 \\
\ThorSimName~B & 5.00 & 2.0 & 5.0 & 40 & 10 & 10 & 0.50 & 2.0 & 30 & 1.0 \\
\ThorSimName~C & 5.00 & 2.0 & 5.0 & 40 & 10 & 10 & 0.50 & 2.0 & 30 & 1.0 \\
\ThorSimName~D & 5.00 & 2.0 & 5.0 & 40 & 10 & 10 & 0.50 & 2.0 & 25 & 1.0 \\
NIHAO-UHD & 8.58 & 2.0 & 5.0 & 40 & 10 & 10 & 0.50 & 3.0 & 25 & 1.5 \\
\hline
\end{tabular}
\tablefoot{Identifier (first column), number of stellar particles (second column), minimum radii for $\rm \Phi_{mid}$ ($r_{min}$, third column), scale parameter for the potential ($a$, fourth column), maximum radii for $\rm \Phi_{mid}$ ($r_{max}$, fifth column), number of summation terms in $\rm \Phi_{mid}$ (sixth and seventh columns), local and global kernel bandwidths (eighth and ninth columns), $X-Y$ range shown in the plots (tenth column), and vertical distance limit (last column).}
\end{table*}
%
%
%
%
%
%
%
\subsection{Model of the potential}
\label{Sect_PotFit}
\par In order to model the axisymmetric component of the gravitational potential, we set the centre of mass of the main galaxy of each simulation at the origin, and rotated the whole system to align the total angular momentum with the vertical axis. For those galaxies that exhibit clumps of accretions in their disc and halo, we manually removed them as their self-gravity constitutes a significant fraction of the potential at that position. By excluding these objects, we avoided the inclusion of misleading local potential values in the fitting process, which are not representative of the global main galaxy's potential.
\par We considered three intervals in galactocentric distance, separated at $r = r_{\text{min}}$ and $r = r_{\text{max}}$, and performed individual fits of the potential in each of them. The primary fit was conducted in the intermediate regime $r_{\text{min}} < r < r_{\text{max}}$, where most of the features are expected. For this interval, we adopted the series proposed by \citet{Hernquist92} as fit of our potential $\Phi_{\rm mid}$, where
\begin{equation}
    \label{Potfit_mid}
    \Phi_{\rm mid}(r,\theta)=\sum_{n=0}^N\sum_{\substack{\ell=0\\ even}}^{L}A_{n,\ell}\Phi_{n,{\ell}}(r,\theta), 
\end{equation}
in which
\begin{equation}
    \label{Phinl}
    \Phi_{n,{\ell}}(r,\theta) = -\sqrt{2{\ell}+1}\frac{(a r)^{\ell}}{(a+r)^{2{\ell}+1}}C_{n}^{(2{\ell}+3/2)}(\xi)P_{{\ell},0}(\cos{\theta}), 
\end{equation}
where $\xi=(r-a)/(r+a)$, $P_{{\ell},0}$ are the Legendre polynomials, and $C_n^{(\alpha)}$ denotes the Gegenbauer polynomials \citep{Gegenbauer1884, Abramowitz1988}. The parameter $a$ is a characteristic radial scale length of the disc \citep{McMillan17}, with values in the range of a few kiloparsecs ($\sim 1-5$ kpc), to improve the convergence of the series in Eq.~\ref{Potfit_mid}. The restriction to even values of $\ell$ in Eq.~\ref{Potfit_mid} imposes a vertically symmetric fit between both hemispheres, while the value $m=0$ in the associated Legendre polynomial implies no azimuthal variations of $\Phi_{\rm mid}$ (hence no dependence on the angle $\phi$). The coefficients $A_{n,\ell}$ were computed by a least-squared fit of the potential value reported by the simulations, performed over a subsample of randomly selected particles since considering the whole simulation in this procedure is computationally expensive.
\par Once $\Phi_{\rm mid}$ is obtained, approximations in the inner ($r\leq r_{\rm min}$) and outer ($r\geq r_{\rm max}$) region were linked to it. For the inner region, we found good fits for the potential assuming the polynomial form
\begin{equation}
    \label{Potfit_in}
    \begin{aligned}
        \Phi_{\rm in}(r,\theta) &= \Phi_{\rm mid}(r_{\rm min},\theta) \cdot \Big[ 1 + a_{00}u + a_{12}u\cos^2{\theta} \\
        &+ a_{22}u^2 \cos^2{\theta} + a_{32}u^3 \cos^2{\theta} + a_{24}u^2 \cos^4{\theta} \Big],
    \end{aligned}
\end{equation}
where $u=1-r/r_{\rm min}$ and the coefficients $a_{ij}$ are the free parameters set by the least-square minimisation. For the outer region, we considered a power-law like decay of the potential with distance,
\begin{equation}
    \label{Potfit_out}
    \begin{aligned}
        \frac{\Phi_{\rm out}(r,\theta)}{\Phi_{\rm mid}(r_{\rm max},\theta)} &= \left[ 1 + \frac{(r-r_{\rm max})^2}{R_{c}^2}\left(1 + \left(\frac{1-q^2}{q^2}\right)\cos^2{\theta}\right) \right]^{-\gamma/2},
    \end{aligned}
\end{equation}
where the free parameters are the characteristic radii $R_c$, the flattening term $q$, and the index of the power law $\gamma$ such that $\Phi_{\rm out}\sim r^{-\gamma}$ at large distances. Note that both $\Phi_{\rm in}$ and $\Phi_{\rm out}$ are continuous with $\Phi_{\rm mid}$ at $r=r_{\rm min}$ and $r=\rm r_{max}$, respectively. The final model of the potential, $\Phi_{\rm fit}$, is obtained by combining the fits across the three distance regimes. 
\par
\begin{figure}[!htbp]
\centering
\includegraphics[width=0.435\textwidth]{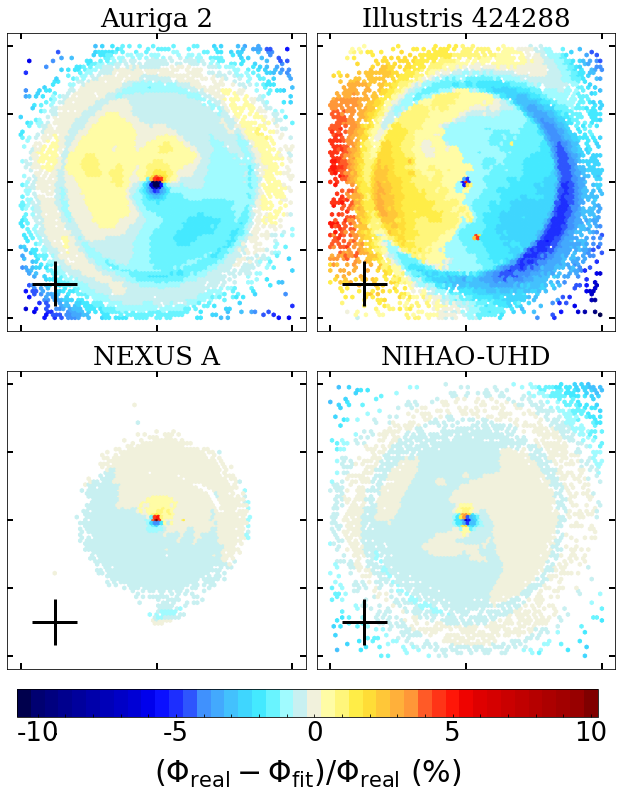}
\caption{Relative residuals of the potential fits (in percentages) for one simulation from each group considered in this work. The cells show the most extreme relative discrepancy between the model and the nominal potential reported in the simulation within the range $|Z|<Z_{lim}$. The black cross in the lower-left corner indicates the scale of $\pm 10$~kpc. The corresponding maps for the other simulations can be found in Figure~\ref{Fig_ResPot}.}
\label{Fig_ResPotPrev}
\end{figure}

 The quality of the potential modelling is illustrated in Figure \ref{Fig_ResPotPrev} for a galaxy from each collaboration group, while Figure \ref{Fig_ResPot} shows the corresponding results for the remaining simulations considered in this work. The colour code indicates the most discrepant relative difference between the reported potential and $\Phi_{\rm fit}$. With the exception of the galactic centre, most regions of the galaxies show good agreement between the model and the original potential, with values of $|\Phi_{\rm real}-\Phi_{\rm fit}| < 0.1 |\Phi_{\rm real}|$, and generally below $\sim 2-3\%$.

%
%
%
%
%
\subsection{Model of the accelerations}
\label{Sec_AccelField}
\par The NewHorizon simulation does not directly provide the potential of the stellar particles; instead, it provides their accelerations, $\vec{a}$, which relate to the potential via $\vec{a} = -\nabla \Phi$. If we assume the potential is axisymmetric, the azimuthal component of the acceleration, $a_{\rm\phi}$, must be zero. This leaves two components of the acceleration to fit: the radial component, $a_{\rm R}$, and the vertical component, $a_{\rm Z}$. Since a single component of the acceleration does not uniquely determine the potential, we need to account for both $a_{\rm R}$ and $a_{\rm Z}$ to estimate the proper $\Phi(r,\theta)$. Therefore, we assumed a model of the potential given by Eq. \ref{Potfit_mid} and tuned the $A_{nl}$ parameters to minimise
\begin{equation}
    \label{Eq_error_forces}
    S^2 = \sum_{i=1}^{N_P} \left(a_{\rm{R},i}+\left. \dfrac{\partial \Phi_{\rm mid}}{\partial R}\right|_{r_i,\theta_i} \right)^2 + \sum_{i=1}^{N_P} \left(a_{\rm{Z},i}+\left. \dfrac{\partial \Phi_{\rm mid}}{\partial Z}\right|_{r_i,\theta_i} \right)^2,
\end{equation}
where the summation is performed over $N_P$ simulation particles and $(R,Z)=(r \sin{\theta}, r\cos{\theta})$ are the cylindrical coordinates. We noted, however, the convergence of Eq. \ref{Eq_error_forces} is quite slow, especially at the galactic plane \citep{Kuijken1995, Dehnen1998}, so we have added the correction term
\begin{equation}
    \label{Eq_correction}
    \Delta\Phi_{\rm mid}(R,Z) = \sum_{n=1}^3 \frac{c_n}{R^n} + c_4\log{R} + c_5 \exp{\left(-\frac{R}{5\rm\ kpc}-\frac{|Z|}{|c_6|}\right)},
\end{equation}
in which the free parameters $c_n$ improved the least-square minimisation of $S^2$. We also tested a correction similar to that used by \citet{Dehnen1998} and implemented in \textsc{galpy} \citep{Galpy} to accelerate the spherical harmonic decomposition; however, Eq. \ref{Eq_correction} led to lower $S^2$ values. 
\begin{figure}[hbtp]
\centering
\includegraphics[width=0.47\textwidth]{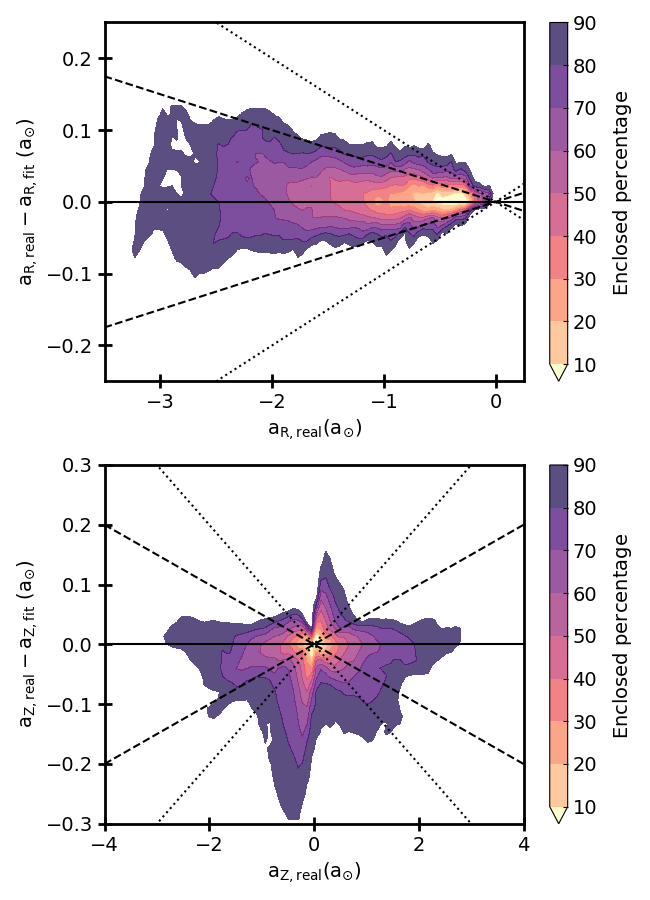}
\caption{Model of the accelerations of the NewHorizon simulation. Upper (lower) panel: discrepancy between the reported radial (vertical) acceleration and its fit as a function of the acceleration itself. The colour-bars indicate the percentage of stellar particles enclosed in each region. Dashed (dotted) lines represent relative errors of $\pm5\%$ ($\pm10\%$). All the accelerations are expressed in units of $a_{\odot}\approx 6895.6~{\rm km^2 s^{-2} kpc^{-1}}$.}
\label{Fig_ResAcc}
\end{figure}
\par The resulting fit of the accelerations is illustrated in Fig. \ref{Fig_ResAcc}. As can be seen, the relative errors in the radial acceleration are mostly smaller than 5\%, while for $a_Z$ these discrepancy can be larger than $\pm 10\%$. We should note, however, this mismatch is appreciated only at low vertical accelerations (e.g., close to the galactic plane), where the radial forces dominate the dynamics. 
\par For the particular case of the NewHorizon simulation, since no improvement in the fit of $\vec{a}$ was observed when $\Phi_{\rm in}$ was considered (see Eq. \ref{Potfit_in}), we decided to retain $\Phi_{\rm mid}$ to model the inner galaxy. On the contrary, a potential of the form $\Phi_{\rm out}$ is used at distances $r > 40$~kpc.
%
%
%
%
%
%
\subsection{Spiral arm overdensities}
\label{Sec_overdens}
\par In order to identify the spiral arms, we applied a bivariate kernel density estimator (KDE) technique similar to that exploited by \citet{Poggio21}. Using this approach, at the location $(X,Y)$ we approximate the local surface density, $\Sigma(X,Y;h)$, as
\begin{equation}
\label{Eq_sigmaKDE}
\Sigma(X,Y;h) = \sum_{i} m_i K(x_i-X;h) K(y_i-Y;h),
\end{equation}
where $K(x;h)\sim (1-x^2/h^2)$ if $|x|<h$ is the Epanechnikov kernel function, which weights the contribution of a particle at distance $|x|$ to the density at origin, assuming a characteristic bandwidth $h$ \citep{Epanechnikov69}. For distances $|x|>h$, the Epanechnikov kernel is zero. The summation in Eq. \ref{Eq_sigmaKDE} is performed over all the particles located at ($x_i, y_i$) with mass $m_i$. Analogously, the global surface density, $\Sigma(X,Y;H)$, is estimated as its local counterpart by using a bandwidth $H>h$, so that the contrast density, $\delta_{\Sigma}$, is determined by their relative difference as
\begin{equation}
\label{Eq_deltaKDE}
\delta_{\Sigma}(X,Y) = \frac{\Sigma(X,Y;h)-\Sigma(X,Y;H)}{\Sigma(X,Y;H)}.
\end{equation}
The selected values for $h$ and $H$ for each simulation are summarised in Table \ref{tab_simparam}. The parameter $h$ is set to approximately match the width of the spiral arms, whereas $H$ represents the typical range of large-scale density variations in the disc. 
\par {Figures \ref{Fig_Actions0}, \ref{Fig_Actions1} to \ref{Fig_ActionsLast} }illustrate the maps of $\delta_{\Sigma}(X,Y)$ for each simulation in the region $|Z|<Z_{\rm lim}$ (values in Table \ref{tab_simparam}). In most simulations, the distribution of $\delta_\Sigma$ is influenced by short-scale distortions, particularly in the outer regions of the galactic disc, where the effects of Poissonian noise become more pronounced due to the low density in these areas. We reduced the contribution of such fluctuations by reconstructing $\delta_{\Sigma}(X,Y)$ from its truncated Fourier series approximation as
\begin{equation}
\label{Eq_deltaKDEFourier}
\tilde{\delta}_{\Sigma}(X,Y) = \sum_{n=0}^{11} \sum_{m=0}^{11} C_{n,m} \exp{\left(\frac{2\pi i}{L_{\rm kpc}}\left[n\cdot X+m\cdot Y\right]\right)},
\end{equation}
where $C_{n,m}$ are the coefficients of the Fourier expansion of $\delta_{\Sigma}$ and $L_{\rm kpc}$ denotes the limits of the square maps shown in Figs. \ref{Fig_Actions0}, \ref{Fig_Actions1} to \ref{Fig_ActionsLast}. Thus, by approximating $\delta_{\Sigma}$ by $\tilde{\delta}_{\Sigma}$ (second columns in Figs. \ref{Fig_Actions0}, \ref{Fig_Actions1} to \ref{Fig_ActionsLast}), we are able to reduce the noise and enhance the larger scale structures in the contrast density maps.
\par By observing the maps of $\delta_{\Sigma}$ and $\tilde{\delta}_{\Sigma}$, we visually identified the tentative trace of the spiral arms (dotted lines in the Figures) for sake of comparison with the features in $J_R$. {The criteria for considering these features as part of the spiral arms depend on: (i) their extent, where segments significantly larger than their width are readily accepted; (ii) its location, as features in the less dense outer regions require special attention due to the Poissonian noise in the $\delta_{\Sigma}$ maps or from artefacts caused by the periodic boundaries of $\tilde{\delta}_{\Sigma}(X,Y)$, a natural consequence of the Fourier reconstruction; and (iii) the degree of agreement between $\delta_{\Sigma}$ and $\tilde{\delta}_{\Sigma}$, where, if the previous criteria are not enough to accept a feature, we check whether its inclusion is consistent with the extension of other nearby structures. Since these criteria rely on visual inspection of the reported maps, some minor or ambiguous features may have been omitted. However, none of these omissions are expected to significantly affect the overall spiral pattern of the galaxy, which is dominated by the large, clearly discernible structures usually located at $R<L_{kpc}$. }
%
%
%
%
%
%
\subsection{Computation of the radial actions}
\par For each simulation, we estimated the orbital parameters, actions and energy by using the potential $\Phi_{\rm fit}$ (see Section \ref{Sect_PotFit}) in the code used in \citet{Palicio2023} instead of the Galactic potential of \citet{McMillan17}. Assuming an axisymmetric potential, this code implements the Stäckel-fudge approximation method \citep[][but see also Appendix B of \citealt{Palicio2023}]{Famaey03,Binney12, Sanders16, Mackereth18, BlandHawthorn19, hadden2024actionangle} to estimate the apocenter, pericenter, eccentricity, maximum galactic height and actions from the input position and kinematics.
\par The maps of the radial action, $J_R$, can be seen in the rightmost columns of Figs. \ref{Fig_Actions0}, \ref{Fig_Actions1} to \ref{Fig_ActionsLast}.  In contrast to \citet{Palicio2023}, where the number of Gaia DR3 stars exceeded that of particles in our simulations, we had to consider an Epanechnikov kernel with $h=0.5$~kpc to effectively increase the number of particles per cell\footnote{For comparison, each cell in Figs. \ref{Fig_Actions0}, \ref{Fig_Actions1}-\ref{Fig_ActionsLast} has $\sim 0.375$~kpc width.} in the maps of $J_R$. The contribution of each particle to the median $J_R$ within a cell is weighted by its stellar mass.
%
%
%
%
%
%
\FloatBarrier
\section{Results and discussions}
\label{Sect_Results}
\subsection{Radial action of the spiral arms}
\label{SubSect_Dynamics}
\par
\begin{figure*}[!htbp]
\centering
\includegraphics[height=0.203\textheight]{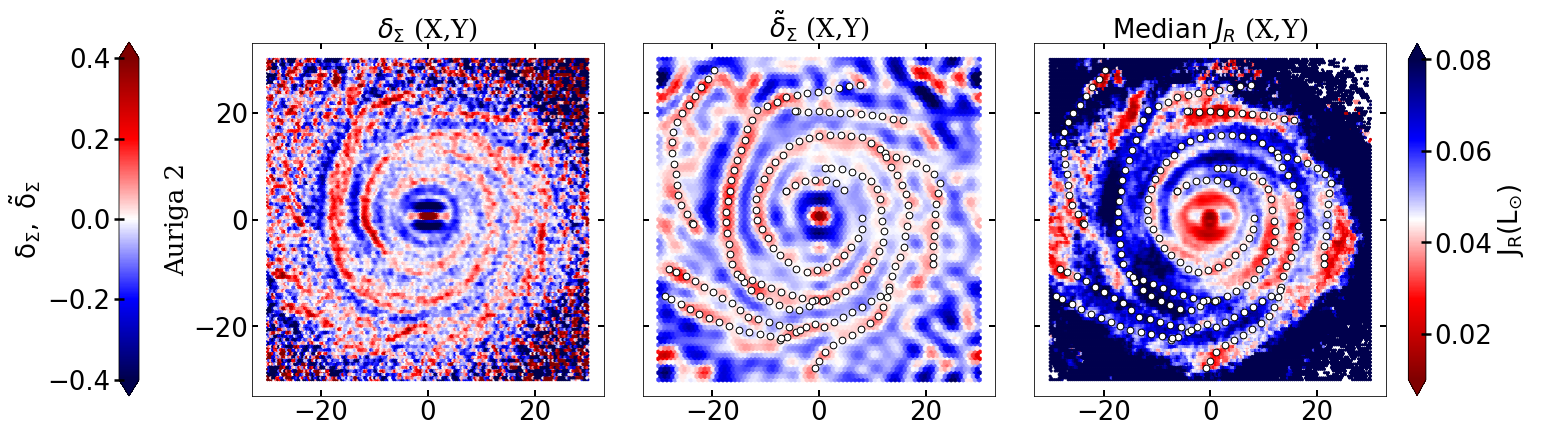}
\caption{Maps of the density contrast (first column), its bidimensional Fourier approximation (second column), and mass-weighted median radial action (third column) for {the \textsc{Auriga 2} simulation}. Open circles in second and third {panels} denote the inferred tracks for the spiral arms. {The corresponding maps for the rest of simulations considered in this work can be found in Figs. \ref{Fig_Actions1} to \ref{Fig_ActionsLast}.}}
\label{Fig_Actions0}
\end{figure*}

 The maps represented in the first columns of Figures \ref{Fig_Actions0}, \ref{Fig_Actions1}-\ref{Fig_ActionsLast} illustrate the contrast of stellar mass density, $\delta_\Sigma$, among the different regions of {each galaxy. In these panels,} we can discern several arc-shaped structures whose densities exceed those of their surroundings (red areas), which we identify as the loci of the spiral arms. They are oriented around the galactic centre, with some bifurcating (as those in \textsc{Auriga 2} and \textsc{Auriga 3}, for example), having kicks in their pitch angles (e.g., \textsc{Auriga 16} and \textsc{Auriga 24}) or showing distortions due to remnants of accreted satellites (e.g.,  \textsc{Illustris 532760}). The contrast density maps filtered for short-scale features, $\tilde{\delta}_{\Sigma}$, are shown in the middle columns of Figs. \ref{Fig_Actions0}, \ref{Fig_Actions1} to \ref{Fig_ActionsLast}, and are adopted as a visual reference for exploring the noisy regions, in which we traced, by visual inspection, those over-density structures that we attributed to the spiral arms (dotted curves in Figs. \ref{Fig_Actions0}, \ref{Fig_Actions1} to \ref{Fig_ActionsLast}). Conversely, we excluded the inner galaxy from our analysis since the high mass of the bulge biases the $\delta_{\Sigma}$ estimator there, leading to unrealistic values of the density contrast (blue annular regions at origin in left and middle columns of Figs. \ref{Fig_Actions0}, \ref{Fig_Actions1} to \ref{Fig_ActionsLast}). Similarly, we did not consider as representative of the main galaxy morphology the features in ${\delta}_{\Sigma}$ created by remnants of accreted satellites, as those seen, for example, in \textsc{Illustris 424288} at (X,Y)$\approx$(-24.5, 9.0), (5.1, -25.0), and (21.0, 17.0)~kpc.
\par The spatial distribution of the median radial action (right columns) also reveals arc-shape structures that trace the low $J_R$ regions. In contrast to those in $\delta_\Sigma$, whose width typically ranges between $\sim 1.5$ and $\sim 2.5$~kpc, the features in $J_R$ can be up to $\sim 4-5$~kpc width (e.g., \textsc{Auriga 16} and \textsc{Illustris 555287}). By confronting the maps of $\delta_\Sigma$ and median $J_R$, we find the overdensities of stellar mass tend to be located at the low radial action areas (dotted curves in middle and right columns), although noticeable discrepancies are observed both in the external zones (e.g., \textsc{Auriga 24}, \textsc{Illustris 549516}, and \textsc{\ThorSimName A}) and the central regions (e.g., \textsc{Illustris 424288}).
\par To evaluate the quality of correspondence between the high $\delta_\Sigma$ and the low radial action arcs, we examined the value of $J_R$ at the location of the nodes that indicate our estimation of the spiral arm tracks (i.e., the circles that form the dotted curves in Figs. \ref{Fig_Actions0}, \ref{Fig_Actions1}-\ref{Fig_ActionsLast}). The separation between two consecutive nodes is 2~kpc. Subsequently, these nodes are classified into three categories based on the degree of agreement with $J_R$: category `G' (good) for the cases in which the $\delta_\Sigma$-$J_R$ correlation is clear, `F' (fair) if a correspondence exists between the features in $\delta_\Sigma$ and $J_R$ despite some mismatch, and `B' (bad) if no agreement is observed. Specifically, the criteria for category `F' include cases that would align after shifts of $\lesssim 1$~kpc in their position, a natural extension of their arcs, or are located in a slightly elevated $J_R$ area (light bluish areas in the map of the median $J_R$). As an illustrative example of this intermediate category, in \textsc{Illustris 424288}, six of the seven last nodes of the spiral arm that ends at ($X$,$Y$)$\approx$(15.3, 0.5)~kpc were classified as `F' nodes. Another example can be found in the nodes between ($X$, $Y$)$\approx$(5.0, 20.), and (10.0, 23)~kpc in \textsc{Illustris 549516}, since their median $J_R$ is slightly above the border value ($0.02 L_{\odot}$) and they bridge the extremes of the arc feature they belong to.
\par {Due to the difference in the width of the arcs seen in the $\delta_\Sigma$ (or $\tilde{\delta}_{\Sigma}$) and $J_R$ mapss, as well as the Poissonian noise in the outer regions (usually at $R\gtrsim L_{kpc}$), a correlation diagram of $\delta_\Sigma$ vs. $J_R$ does not clearly show the connection between the overdensities and the zones of low radial action. However, we present these diagrams in Appendix \ref{app_correlation_diagram} where, despite the large scatter, all the simulations show an anticorrelation between $\delta_\Sigma$ and $J_R$.}
\par The resulting classification revealed the nodes tagged as `G' constitute, on average, the 62.6\% of the total nodes of a simulation. This percentage, however, ranges from 14.7\% for the \textsc{\ThorSimName A} simulation to 88.1\% for \textsc{Auriga 16}, while 18 of the 21 galaxies have percentages for the `G' sample $\gtrsim$50\%\footnote{This includes the \textsc{\ThorSimName D} simulation, whose percentage for the `G' subset is 49.3\%. }. If we consider the combination of categories `G+F' as positive matches, the percentage of agreement rises, on average, to 80.2\%; while all the simulations exceed the 70\% threshold but two: \textsc{\ThorSimName A} (57\%) and \textsc{Auriga 2} (67\%). The percentage of agreement for each galaxy is indicated in Table \ref{tab_tagging}.
\begin{table}[h]
	\centering
	\caption{Statistics of the node classification based on the agreement between the ${\delta}_{\Sigma}$ overdensity and the low $J_R$ regions. }
	\label{tab_tagging}
	\begin{tabular}{lcccc}
\hline
SimID & $\rm N_{nodes}$ & G(\%) & F(\%) & B(\%)\\
\hline
Auriga 2 & 218 & 48.2 & 18.8 & 33.0 \\
Auriga 3 & 156 & 66.0 & 17.3 & 16.7 \\
Auriga 16 & 134 & 88.1 & 5.2 & 6.7 \\
Auriga 19 & 81 & 70.4 & 17.3 & 12.3 \\
Auriga 24 & 163 & 71.8 & 6.7 & 21.5 \\
Auriga 27 & 111 & 72.1 & 14.4 & 13.5\\
Illustris 424288 & 131 & 75.6 & 11.5 & 13.0 \\
Illustris 464163 & 234 & 80.3 & 6.8 & 12.8 \\
Illustris 479290 & 162 & 55.6 & 15.4 & 29.0 \\
Illustris 509091 & 155 & 86.5 & 11.0 & 2.6 \\
Illustris 532760 & 156 & 75.0 & 7.1 & 17.9 \\
Illustris 537941 & 115 & 67.0 & 9.6 & 23.5 \\
Illustris 549516 & 188 & 60.6 & 11.2 & 28.2\\
Illustris 552414 & 142 & 69.0 & 19.7 & 11.3\\
Illustris 555287 & 119 & 73.1 & 16.8 & 10.1\\
NewHorizon & 162 & 72.8 & 8.0 & 19.1 \\
\ThorSimName~A & 300 & 14.7 & 42.7 & 42.7\\
\ThorSimName~B & 282 & 34.4 & 41.5 & 24.1\\
\ThorSimName~C & 273 & 25.3 & 46.5 & 28.2\\
\ThorSimName~D & 268 & 49.3 & 21.6 & 29.1\\
NIHAO-UHD & 97 & 59.8 & 19.6 & 20.6 \\
\hline
\end{tabular}
\tablefoot{Identifier (first column), number of nodes on the track of the spiral arm (second column), percentage of nodes classified in the G, F, and B categories (third, fourth, and fifth columns, respectively).}
\end{table}
%
%
%
%
%
%
%
%
\subsection{Anomalous scenarios}
\label{SubSect_BadCases}
\par Although most of the spiral arms analysed above show a correlation with the low $J_R$ regions, we have detected a few  discrepancies in some cases, especially in the \ThorSimName family of simulations, as well as in other numerical galaxies not included above but described and discussed as follows. In the {first} panel of Fig. \ref{Fig_ActionsLast}, we can observe a misalignment between the spiral arms, as traced by the overdensity in ${\delta}_{\Sigma}$, and the low $J_R$ regions for the \textsc{\ThorSimName A} simulation. This mismatch is more evident at the outermost parts ($R\gtrsim$~20 kpc) of the main spiral arms of \textsc{\ThorSimName A}, \textsc{\ThorSimName C}, and \textsc{\ThorSimName D} {(first, third, and fourth panels in Fig. \ref{Fig_ActionsLast}, respectively)}; as well as in the innermost $\sim$10~kpc where the spirals wind up bringing closer the features in $J_R$. We interpreted these mismatches as a consequence of a configuration far from the equilibrium for these simulations, as evidenced by the void in the inter-arm region at $Y\gtrsim$20~kpc. Furthermore, we verified the radial velocity field, $V_R$, suggests an eventual wind up of the major arm there, removing the gap and leading to a more relaxed configuration. The same argument applies in the $R\lesssim 10$~kpc regions, where the particles in the arm (inter-arm) show inwards (outwards) radial motions. In other words, the spatial shifts observed between the high density and the low $J_R$ areas are compatible with a spiral pattern that is still winding up.
\par Other major discrepancy concerns the spiral arm detected at the lowermost boundary of the above mentioned gap; such as that starting at $(X,Y)\approx (28, 0)$~kpc in \textsc{\ThorSimName A} but with analogous examples in all the \textsc{\ThorSimName} simulations. This spiral arms do not lie on a low radial action area but trace the envelope of the highest $J_R$ region. Although the out-of-equilibrium scenario described above can explain such discrepancy with the other simulations, there is an alternative explanation for such mismatch concerning the kernel technique described in Sec. \ref{Sec_overdens}: the abrupt transition from moderately populated (i.e., region below the dotted curve starting at X=20~kpc, Y=0~kpc) to almost empty areas (above the dotted curve) can produce misleading estimations of the density contrasts, which might be misinterpreted as spiral arms. This idea is explained in detail in Appendix \ref{Sec_Discontinuity}.
\par
\begin{figure*}[htbp]
\centering
\includegraphics[height=0.375\textwidth]{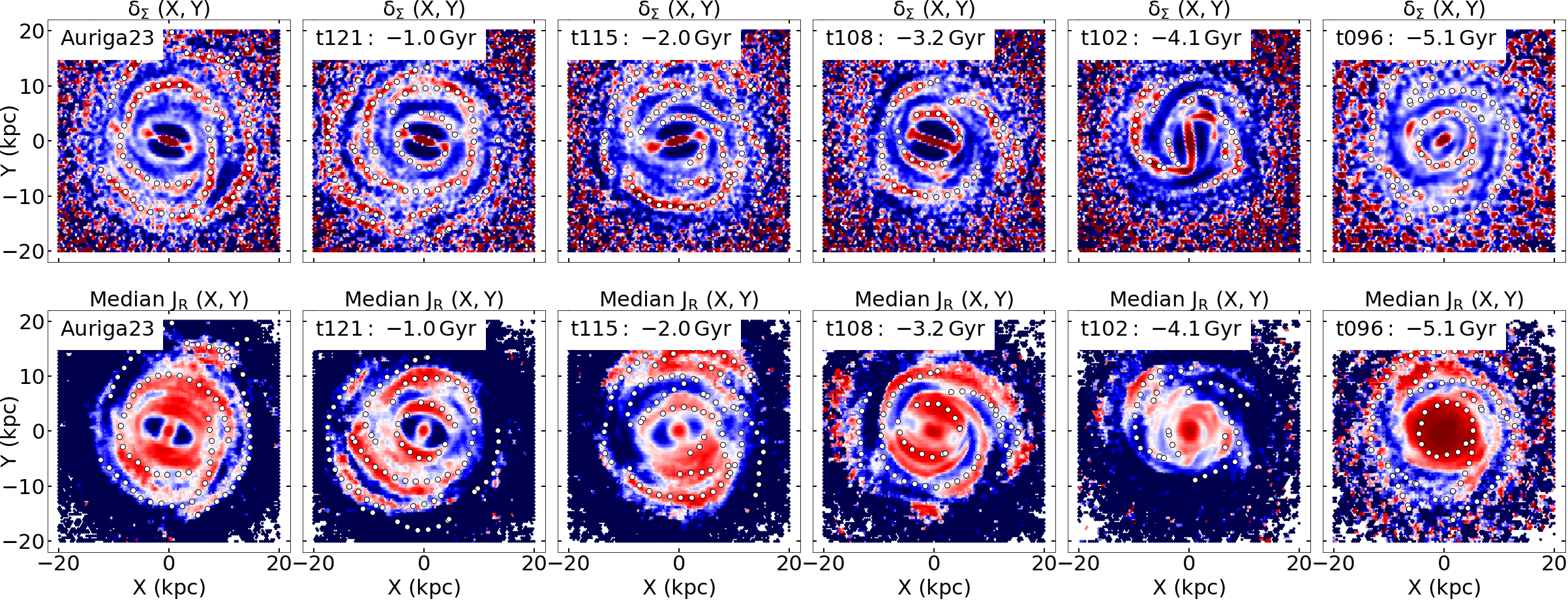}
\caption{Maps of the density contrast (upper panels) and mass-weighted median radial action (bottom panels) for the \textsc{Auriga 23} barred simulation analysed in Section \ref{SubSect_BadCases}. Open circles denote the inferred tracks for the spiral arms.}
\label{Fig_Action23}
\end{figure*}

 In addition to the \textsc{Auriga} simulations presented in Figs. \ref{Fig_Actions0} and \ref{Fig_Actions1}, we analysed the peculiar case of \textsc{Auriga 23}, whose overdensity map (upper-left panel in Fig. \ref{Fig_Action23}) suggests the presence of a long bar in the central regions. As commonly done in the literature \citep[e.g.,][]{MartinezValpuesta2017, Thomas2023, BlandHawthorn2024}, we used the ratio between the amplitudes of the dipolar $A_2$ and the axisymmetric $A_0$ components of the surface density to confirm the existence of the bar and quantify its strength. We computed $A_2/A_0$ over the range $R=0.5$ to 7.5~kpc using bins of $\Delta R=0.5$~kpc width with a 50\% overlap. 
\par
\begin{figure*}[htb]
\centering
\includegraphics[width=0.95\textwidth]{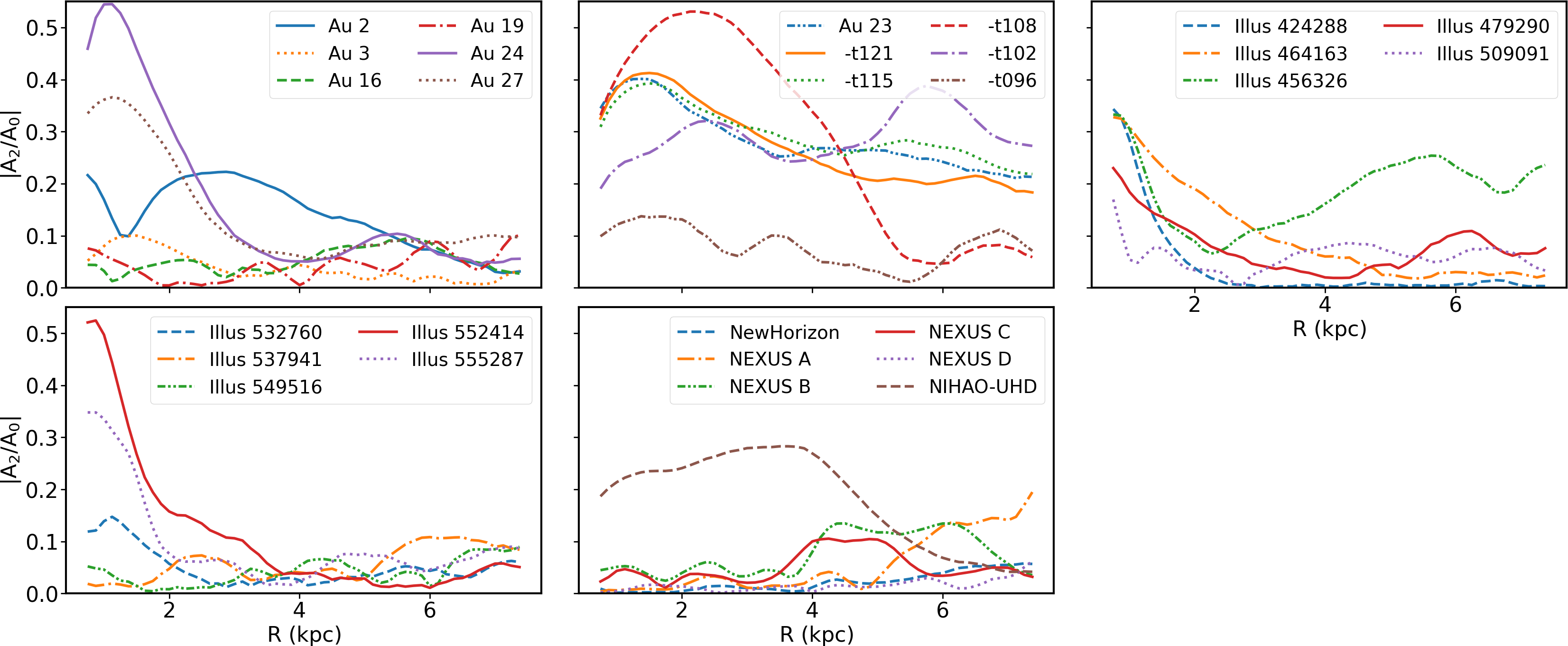}
\caption{Absolute value of the dipolar-to-axisymmetric component of the density ratio, $|A_2/A_0|$, for all the simulations considered in this work. Large values of $A_2/A_0$ denote great contribution from elongated structures, like a bar, to the local density. For the sake of visualisation, the complete set of simulations has been divided into five panels.}
\label{Fig_BarAmplitude}
\end{figure*}

 As shown in Fig. \ref{Fig_BarAmplitude}, \textsc{Auriga 23} exhibits a large $A_2/A_0$ ratio ($\sim 0.4$) compared to most of the simulations, surpassed only by \textsc{Auriga 24} ($A_2/A_0 \approx 0.54$), two earlier snapshots of \textsc{Auriga 23} from $\sim 3$~Gyr and $\sim 1$~Gyr ago ($A_2/A_0 \approx 0.53$ and $\approx 0.41$, respectively), and \textsc{Illustris 552414} ($A_2/A_0 \approx 0.52$) with a $\sim 2$~kpc half-length bar. Despite being stronger, the bars in \textsc{Auriga 24} and \textsc{Illustris 552414} {do not seem} to cause significant distortions in the distribution of radial actions. {To explore this in detail, we located the resonances associated with the $m=2$ and $m=4$ perturbers of these galaxies and checked whether the radial action maps show any imprint of them at these positions. In particular, when two perturbers with different pattern speeds coexist, an overlap of their resonances can induce a significant exchange of angular momentum \citep{Quillen03,Minchev10, Minchev11} in a non-linear fashion (i.e., the effect of the combined perturbers exceeds that produced by each of them individually)---especially if the bar's corotation or outer Lindblad resonance is involved. This change in angular momentum would imply a variation in the radial action whenever a resonance other than corotation participates \citep{Sellwood2002}, as suggested by the increase in the radial velocity dispersion reported by \citet{Minchev10}. }
\par {Therefore, for the \textsc{Auriga 24} and \textsc{Illustris 552414} galaxies, we located the resonances of their $m=2$ and $m=4$ perturbers, estimating their pattern speeds through the 2D Fourier mode velocity method \citep{Pfenniger23}. We noted that, due to the complexity of the structures in the discs of simulated galaxies—where multiple perturbers can coexist with different pattern speeds and transient behaviour—the inferred rotation $\Omega$ and epicyclic $\kappa$ frequencies may not decrease monotonically with radius. Consequently, a galaxy may exhibit more than one outer Lindblad resonance, as is the case for \textsc{Auriga 24} and \textsc{Illustris 552414}. }
\par
\begin{figure}[htb]
\centering
\includegraphics[width=0.48\textwidth]{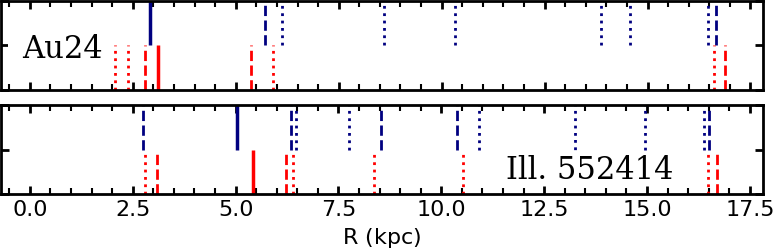}
\caption{{Location of the resonances created by the $m=2$ and the $m=4$ perturbers in the \textsc{Auriga 24} and \textsc{Illustris 552414} simulations. Blue (red) vertical lines refer to the $m=2$ ($m=4$) perturber; while the linestyle denotes the frequency ratio involved: corotation (solid lines), first harmonics $2:1_b$ and $4:1_s$ (dashed lines), second harmonics $1:1_b$ and $2:1_s$ (dotted lines).}}
\label{Fig_Resonances}
\end{figure}

 {As Figure \ref{Fig_Resonances} illustrates, the \textsc{Auriga 24} simulation shows three resonance overlaps created by the $m=2$ and $m=4$ modes (hereafter denoted by the subscripts `b' and `s', respectively): an overlap between the corotation resonances at $R_{CR,b}\approx R_{CR,s}\approx 3.0$~kpc, and two overlaps involving the $2:1_b$, $2:1_s$, and $4:1_s$ resonances at $\sim 5.75$~kpc and $16.5$~kpc. Among them, only the corotation overlap appears to have a significant impact on the radial action distribution, acting as a barrier for the innermost low-$J_R$ regions. }
\par Conversely, in the \textsc{Illustris 552414} simulation, the corotation radii $R_{CR,b}$ and $R_{CR,s}$ are slightly more separated, while the $2:1_b$ resonances overlap with the ultraharmonic $2:1_s$ and the outer Lindblad resonances $4:1_s$ at $5.5$, $8.5$, $10.5$, and $16.5$~kpc. However, we detected no distinct features in the $J_R$ distribution that suggest any influence from these overlaps, except for limiting the innermost low-$J_R$ regions at the corotation radii, similar to \textsc{Auriga 24}. In contrast to \textsc{Auriga 24} and \textsc{Illustris 552414}, the \textsc{Auriga 23} snapshots reveal two crescent-shaped regions of low $J_R$ located on either side of the bar's minor axis, indicating a notable impact on the radial action distribution {in this area}.
\par
\begin{figure*}[htb]
\centering
\includegraphics[height=0.375\textwidth]{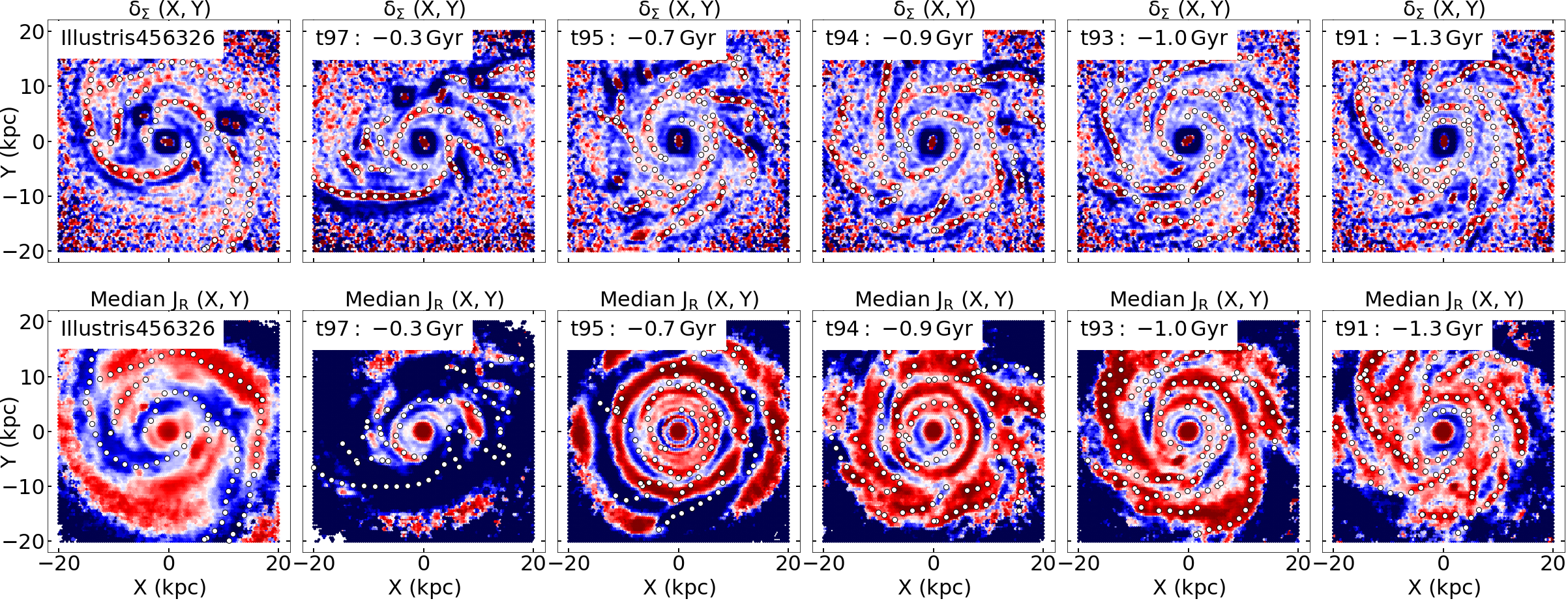}
\caption{Maps of the density contrast (upper panels) and mass-weighted median radial action (bottom panels) for different snapshots of the \textsc{Illustris 456326} simulation, which is affected by a recent merger, as analysed in Section \ref{SubSect_BadCases}. Open circles denote the inferred tracks of the spiral arms.}
\label{Fig_Merger}
\end{figure*}

 Finally, the analysis of the \textsc{Illustris 456326} simulation revealed a significant mismatch between the overdensities in ${\delta}_{\Sigma}$ and the features in the median $J_R$ (Fig. \ref{Fig_Merger}). The latter exhibited very extended structures compared to those observed in the other simulations. To evaluate the reasons behind these discrepancies, we explored the recent history of this simulation using earlier snapshots. 1.35~Gyr ago, the spiral arms of \textsc{Illustris 456326}t91 showed a strong correlation with the low-$J_R$ regions, representing one of the clearest examples in our study. This agreement persisted until snapshot \textsc{Illustris 456326}t94, corresponding to 0.9~Gyr ago, where a ring-shaped area of low radial action emerged at $R \approx 6.5$~kpc, and slight misalignments between the spiral arms and the arcs in $J_R$ became noticeable. 
\par At later times, the maps from subsequent snapshots reveal a new configuration of the overdensities in ${\delta}_{\Sigma}$, which reorganised to eventually form two large spiral arms. Meanwhile, the ring-shaped low-$J_R$ area expanded to $R \approx 10.8$~kpc in snapshot \textsc{Illustris 456326}t95 (0.7~Gyr ago). In the following snapshots, there is a complete misalignment between the overdensities and the arcs in $J_R$, with no discernible patterns emerging. At the present time, the \textsc{Illustris 456326} simulation shows two banana-shaped regions characterised by low radial action, of which only the uppermost region includes a spiral arm.
\par We interpreted the evolution described above as the result of an interaction between the main \textsc{Illustris 456326} galaxy and a nearby satellite, whose mass comprises 20\% of the total system's mass. The pericenter of this encounter was located, at most, at 44.2~kpc (snapshot \textsc{Illustris 456326}t95) from the centre of the main body, producing an overlap between the discs of both galaxies approximately $\sim 0.7$~Gyr ago. The effects of this interaction can be observed in Fig. \ref{Fig_ResPotMerger}, where the companion caused significant distortions and asymmetries in the potential of the main galaxy.
%
%
%
%
%
%
%
%
%
%
%
%
%
\subsection{Age of the spiral arm population}
\label{SubSect_Ages}
\par Apart from the low $J_R$-overdensity correspondence, the results of \citet{Palicio2023} suggested the spiral arms 
might be also supported by populations older than expected. In that work, we were able to discern the spiral arms in the distribution of $J_R$ using a photometrically selected sample of Gaia DR3 giant stars. Furthermore, by reproducing the KDE technique described in Sect. \ref{Sec_overdens}, we observed overdensities of giant stars at the locations of the spiral arms reported by \citet{Poggio21} and traced by upper main sequence stars. The idea of old populations contributing to the spiral arm structure is not new but was originally noted by \citet{Zwicky1955} and \citet{Schweizer76}, although for `grand design' spirals \citep{Eskridge2002}. However, the spiral arms of the Milky Way are thought to be intermediate-scale or flocculent types \citep{Binney2008, Quillen2018, MartinezMedina22}, and they are not expected to be long-lived structures \citep{Oort1970, Elmegreen84}.
\par The methodology of \citet{Palicio2023}, however, offers an alternative explanation for the features seen in their $J_R$ maps with giants stars: their uncertainties in the photometric selection criteria (observational errors, incorrect estimation of the extinction, etc...) may have misclassified some young sources as giants, in which the former would be preferentially located at the spiral arms regions they were born. On the contrary, the giant population would show a more uniform spatial distribution since it is older. Although the polluting young sample were smaller in volume compared to the old one, the KDE method would enhance the spatial asymmetries of the former revealing the spiral arms.
\par
\begin{figure*}[btbp]
\centering
\includegraphics[height=0.17\textheight]{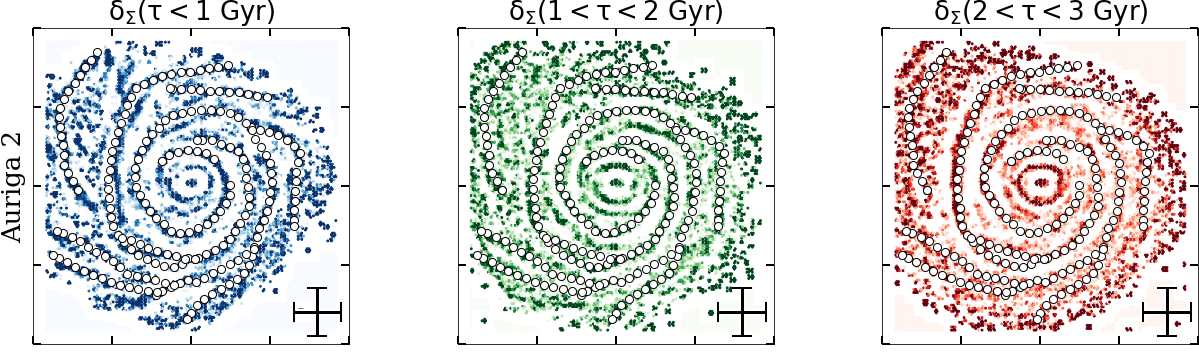}
\caption{Density contrast maps of the \textsc{Auriga 2} simulation for three distinct age intervals: stellar particles younger than 1 Gyr (left column), those aged between 1 and 2 Gyr (middle column), and those between 2 and 3 Gyr (right column). Open circles indicate the spiral arm tracks inferred from Fig. \ref{Fig_Actions0}. Cross in the bottom-right corner indicates the scale of $\pm5$~kpc.}
\label{Fig_AgeContrastCited0}
\end{figure*}

 In order to address whether the old population can trace the spiral arms or not, we have applied the KDE technique on three subsets of different ranges of age, $\tau$: the young ($\tau\leq 1$~Gyr), intermediate ($1\leq\tau<2$~Gyr) and old ($2\leq\tau<3$~Gyr) populations. The resulting maps for ${\delta}_{\Sigma}$ are illustrated in Figs. \ref{Fig_AgeContrastCited0}, \ref{Fig_AgeContrastNonCited0}-\ref{Fig_AgeContrastNonCitedLast}. In general, all the spiral arms traced by the whole simulation (dotted lines) are observed at the three age intervals, although a few minor discrepancies are found in \textsc{Illustris 464163}, where half of the spiral arm starting at $(X,Y)\approx(1.4, -10.4)$~kpc is not traced by the oldest sample;  in \textsc{Illustris 479290}, where the small overdensities at the upper right corner of the maps can be only identified in the young subset; as well as in \textsc{NewHorizon}, in which the most external spiral arms are not discernible in populations older than 1~Gyr. In contrast, the spiral arms of the NIHAO-UHD simulation are almost exclusively supported by the young population.
\par Our results align with the findings of \citet{Lin2022} and \citet{Uppal23}, who demonstrated that it is possible to trace the Local and Outer spiral arms, respectively, using the red clump population, whose age distribution is expected to peak at $\sim1-4$~Gyr \citep{Girardi01, GirardiRC}. Similarly, \citet{Khanna2024} identified structures in the distribution of red clump stars consistent with the spiral arms traced by \citet{Drimmel2000}, although none correspond to the Local Arm. In a recent study, \citet{Debattista2024} identified the spiral arms of their simulated galaxy by analysing the population older than $2$~Gyr. 
%
%
\par We do not observe, however, any significant misalignment in the spiral arm traces across the three age bins \citep{Rand89}. This contrasts with the findings of \citet{Hao2021}, who reported offsets using masers and young open clusters. A key distinction is that their oldest age interval extends up to $0.1$~Gyr, whereas our youngest age bin is an order of magnitude greater. Consequently, it is likely that our maps lack the resolution necessary to disentangle the offsets between populations of such small age difference.
%
%
%
%
%
%
%
\subsection{Relation to the Milky Way}
\label{SubSect_MW}
\par
\begin{figure}[htbp]
\centering
\includegraphics[width=0.48\textwidth]{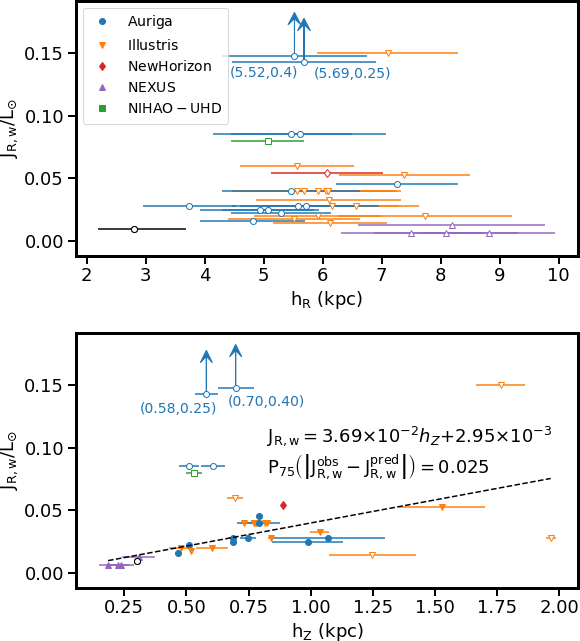}
\caption{Dependence of $J_{R,w}$ with the structure and kinematic parameters of the simulations: radial scale-length (upper panel) and vertical scale-length (lower panel). Coloured markers refer to simulation families: \textsc{Auriga} (blue circles), \textsc{Illustris} (orange triangles), \textsc{NewHorizon} (red diamond), \textsc{\ThorSimName} (purple triangles), and \textsc{NIHAO-UHD} (green square). Black markers denote measurements for the Milky Way reported in the literature (see main text). Open markers denote outliers excluded from the linear fit (dashed line) by the $\sigma$-clipping algorithm.}
\label{Fig_MWCorrelations}
\end{figure}

 The arc-shaped structures identified in the maps of radial actions are not delimited by a single value of $J_R$ common to all the simulations. Instead, each galaxy requires an individual reference value, $J_{R,w}$, that separates the high and low radial action regions (see rightmost colour-bars in Figs. \ref{Fig_Actions0}, \ref{Fig_Actions1}-\ref{Fig_ActionsLast}). Motivated by the results of \citet{Jia2023} and \citet{jia2024investigatingverticaldistributiondisk}, who identified a correlation between the vertical scale-length, $h_Z$, and the mean radial action along the disc of the form $J_R \sim (h_Z - b)^2$, we investigated a potential connection between $J_{R,w}$ and the vertical and radial ($h_R$) scale-lengths. These parameters are represented on the horizontal axis of Fig.~\ref{Fig_MWCorrelations}.
\par The scale-lengths $h_R$ and $h_Z$ were computed assuming exponential fits for the histograms of stellar masses along the radial and vertical directions, respectively. Since the discs of the analysed simulations show non-negligible deviations from exponential profiles in the radial direction, we considered the average of two estimations for $h_R$: one imposing $R < L_{kpc}$ and another with $R < L_{kpc}/2$ as the maximum galactocentric distance for the histogram. In both cases, the criteria $|Z| < Z_{lim}$ and $0.2L_{kpc} \leq R$ were applied to focus on the galactic plane and exclude the bulge regions, respectively. Similarly, for $h_Z$, we averaged the estimations within $|Z| < Z_{lim}$ and $|Z| < Z_{lim}/2$, imposing $0.2L_{kpc} \leq R < L_{kpc}$. The discrepancy between both estimations is denoted by the (symmetric) error bars in Fig.~\ref{Fig_MWCorrelations}.
\par As shown in the upper panel of Fig.~\ref{Fig_MWCorrelations}, no clear trend $J_{R,w}(h_R)$ can be inferred from the distribution of data points, primarily due to the large errors associated with the estimation of $h_R$. Its vertical counterpart, however, shows a positive correlation between the radial action $J_{R,w}$ and $h_Z$. To confirm this correlation, we performed a linear fit excluding outliers through a $\sigma$-clipping procedure: in each iteration, only the data points whose absolute discrepancy lay within the 75$^{\mathrm{th}}$ percentile participated in the next fit (filled markers in Fig.~\ref{Fig_MWCorrelations}). The resulting parameters of the fit confirmed our visual interpretation of the $J_{R,w}$ vs. $h_Z$ correlation, showing a positive slope of $3.69 \times 10^{-2} L_{\odot}/\rm{kpc}$. This correlation is analogous to that found by \citet{Jia2023} at different ranges of Galactocentric radii, although we adopted a simpler functional form for the fit. \citet{Jia2023} proposed a parabolic fit for $J_{R,w}(h_Z)$, as predicted by some distribution functions for the Milky Way under the epicycle approximation (see Problem 4.21 in \citealt{Binney2008}), while we found that a linear relation provides an adequate fit to the data points. It is important to note, however, that the parameter $J_{R,w}$ does not correspond to the mean radial action as in \citet{Jia2023}, and we did not subdivide the data into radial bins as they did. Consequently, the observed trends of $J_{R,w}$ with scale lengths or velocity dispersions might appear linear rather than quadratic.
%
%
\par For comparison with the Milky Way, we included estimations for $h_R$ and $h_Z$ in the diagrams of Fig.~\ref{Fig_MWCorrelations} (black markers\footnote{The Milky Way data points were not included in the iterative fitting procedure.}). We adopted the value $J_{R,w} = (9.4 \pm 0.6) \times 10^{-3} L_{\odot}$ for the Milky Way from Figure 1 of \citet{Palicio2023}, while those for $h_R$ were extracted from the compendium presented in \citet{BlandHawthorn16}. For the vertical scale length, we considered $h_Z \approx 0.3$~kpc, as widely reported in the literature for decades \citep{Gilmore83, Juric08}. 
\par The large error bars associated with the radial scale-length estimates prevent reliable comparisons, in which the Milky Way's $h_R$ remains the shortest among them. Conversely, the position of the Milky Way in the $J_{R,w}$ vs. $h_Z$ plane aligns with expectations from the linear fit, as well as with those of the \textsc{\ThorSimName} galaxies. 
\par As noted by \citet{Jia2023}, the correlation between $h_Z$ and the radial action can be explained as a consequence of the disc's underlying heating mechanisms, which broaden the distribution of radial actions and increase the velocity dispersions. This implies greater kinetic energy in the vertical direction, allowing stars to reach larger galactic heights. Since radial actions are always positive, the broadening of the $J_R$ distribution is associated with a shift of the mean and median towards higher values. This effect can be visualised assuming an exponential distribution for $J_R$, as in Fig. 3 of \citet{Jia2023}, where the mean and median are proportional to the scale factor. We attribute the lack of correlation between $J_{R,w}$ and $h_R$ to their evolution through different mechanisms: the radial action is a key parameter determining the orbital eccentricity, $ecc$, around the Galactic centre ($ecc \propto \sqrt{J_R}$ under the epicyclic approximation, as shown in Eq. 3.261 of \citealt{Binney2008}). Thus, increasing the radial action implies larger $ecc$, leading to an outward displacement of the apocenter\footnote{The eccentricity is usually approximated as $(r_{apo} - r_{peri})/(r_{apo} + r_{peri})$.}, $r_{apo}$. In this configuration, stars can visit outer disc regions, which might suggest, erroneously, that $h_R$ has increased due to stellar mass being pushed outward. However, larger eccentricities also imply shorter galactocentric distances for the pericenter, resulting in stellar mass transport towards central regions and a reduction of the radial scale-length. Consequently, changes in radial action induced by heating mechanisms are unlikely to significantly impact $h_R$, particularly within the epicyclic approximation regime. However, when this approximation does not hold, as in the case of very eccentric orbits, a star may spend more time at galactocentric distances greater than its guiding radius than within it. This implies, on average, a net outward displacement of mass. In contrast, changes in angular momentum, $L_Z$, directly affect the guiding radii of the orbit, a proxy for the mean galactocentric distance, potentially triggering radial migration that redistributes mass outward, thereby increasing $h_R$. In alignment with this interpretation, \citet{Frankel2020} reported changes in the angular momentum of the Milky Way thin disc that are approximately 10 times larger than those in radial action. In a recent study, \citet{Hamilton2024} explored several models of spiral arms to reproduce the observed heating-to-migration ratio of $\sim 0.1$. They concluded that the non-linear horseshoe transport mechanism \citep{Sellwood2002} is compatible with the observations, albeit under rather specific conditions related to the spatial extent of the perturbation amplitude or the presence of greater pitch angles in the past. However, in the former scenario, the non-horseshoe scattering mechanism can also be consistent with the data.
%
%
%
\par It is interesting to compare the radial action maps presented in this work to the one obtained for the Milky Way, represented in Fig. 1 of \citet{Palicio2023}. In our Galaxy, however, the comparison between the $J_R$ maps and the spiral structure is not trivial, because there is no consensus about the geometry and number of spiral arms in the Galactic disc. Low-$J_R$ features from \citet{Palicio2023} tend to be co-located with the spiral arm geometry proposed by \citet{Reid14}, with the exception of the Scutum arm and a small part of the Perseus arm in the third Galactic quadrant \citep[see second panel of Fig. 4 in ][]{Palicio2023}. The general good agreement between the spiral arm model from \citet{Reid14} and the observed low-$J_R$ features would suggest a connection between the two, as seen in the majority of the simulated galactic discs analysed in this work.
\par However, a larger pitch angle for the Perseus arm has been found by other works based on young stars in Gaia \citep{Zari21,Poggio21,PVPDrimmel} and Cepheids \citep{Drimmel2024}, in agreement with the model by \citet{Levine2006} based on HI observations (see their spiral arm n. 2). Such pitch angle would also be consistent with the map obtained with the giant sample in \cite{Palicio2023}.  Assuming the Perseus arm geometry from these works, the low-$J_R$ feature in the outer disc mapped by would not coincide with a spiral arm \citep[see first and last panels of Fig. 4 in ][]{Palicio2023}. Therefore, in this scenario, the outer disc of the Milky Way would be more similar to the out-of-equilibrium simulated galaxies analysed in this work \citep{Antoja18,McMillan2022,Poggio2024}; although similar discrepancies have also been observed in less disturbed simulations.
\par Based on all spiral arm geometries explored, the structures tentatively associated with the Local and Perseus spiral arms, respectively, merge in an area of low $J_R$, showing a certain misalignment with the literature --- except for the 17th group of Cepheids reported in \citet{Lemasle22}. By analysing the maps for $\delta_{\Sigma}$ and $J_R$ from the simulations, we identified similar structures, where the region between two nearby spiral arms also presents low radial action. This is the case for the bifurcation at $(X,Y)\approx (-8, 0)$~kpc in \textsc{Auriga 19} ({third} row in Fig. \ref{Fig_Actions1}), at $(10, -6.6)$~kpc in \textsc{Auriga 27} ({fifth} row in Fig. \ref{Fig_Actions1}), at $(-5.3, -7.3)$~kpc in \textsc{Illustris 555287} ({fourth} row in Fig. \ref{Fig_Actions3}), and the gap between two spiral arms in \textsc{\ThorSimName D} at $(10.6, 0)$~kpc ({fourth} row in Fig. \ref{Fig_ActionsLast}).  While reaching a definitive conclusion about the Milky Way remains challenging, there is no doubt that mapping $J_R$ features in the Galactic disc provides a unique perspective on the dynamical history of the Galaxy, as demonstrated by the simulations analysed in this work.
%
%
%
%
%
%
%
\section{Conclusions}
\label{Sect_Conclu}
\par Our analysis of a set of simulations from independent groups {supports} the correlation between the location of the spiral arms and the low radial action features, {as proposed by} \citet{Palicio2023} using Gaia eDR3 and DR3 data \citep{GaiaEDR3,GaiaDR3}. {In that work, the features in $J_R$ were mainly consistent with the models of the Sagittarius, Local, and Perseus spiral arms proposed by \citet{Reid14} in the first and second quadrants ($0^{\circ} \leq \ell \leq 180^{\circ}$), while at negative longitudes, no preferential model emerged from the comparison with the literature. For our sample of simulations, such} correspondence was observed in 20 of the 23 spiral galaxies, with an average percentage of agreement ranging from the 64\% to the 88.1\%, depending on the strictness of the goodness criterion. Conversely, for the remaining two simulations, the mismatch between the ${\delta}_{\Sigma}$ and the $J_R$ maps can be explained by an out of equilibrium state, such as that triggered by an interaction with a companion galaxy (\textsc{Illustris 456326}, \textsc{\ThorSimName A-D}), and by border effects of the KDE technique, as these observed in the \textsc{\ThorSimName} simulations and explained in Appendix \ref{Sec_Discontinuity}.
\par Although restricted to the particular cases of \textsc{Auriga 23} and \textsc{NIHAO-UHD}, we found no evidence of a great influence of the bar on the distribution of the radial actions at the locations of the spiral arms. We must exclude from this observation, however, the central regions in which the bar is confined, as two arc-shaped areas of low $J_R$ have emerged near the minor axis. In any case, all the conclusions derived for the barred galaxy scenario should be interpreted with caveat since our methodology assumes the potential is predominantly axisymmetric. Therefore, it might result unrealistic if the bar contributes significantly to the total potential, requiring more advanced approaches for the action estimation, such as that proposed by \citet{Debattista2024}.
\par On the other hand, strong interactions with satellites can destroy the existing spiral arm pattern, leading to a new configuration in which the emerging spiral arms do not lie on low $J_R$ areas, as observed in the recent history of the \textsc{Illustris 456326} simulation. This perturbed scenario is expected to be transitory and eventually evolve to a new equilibrium state in which the $J_R$-${\delta}_{\Sigma}$ anticorrelation is recovered. Further re-simulations of post-interaction galaxies are required for confirming this hypothesis. 
\par The maps of the density contrast, ${\delta}_{\Sigma}$, computed for different bins in age revealed the population older than $1$~Gyr can trace the spiral arms. Furthermore, at the spatial resolution of the considered simulations, no significant misalignment among the spiral arm tracks has been observed when compared the three age bins. This contradicts the classical picture of the spiral arms traced by the young sources recently formed in, implying differences in the mean age between the leading and trailing sides of the arms. We recall, however, this contrast in age might not be discernible in our maps due to the spatial resolution. Addressing whether the old population was formed in the spiral arms they are observed in or have been trapped later is beyond the scope of this paper.
\par The comparison of the limiting value for the low $J_R$ regions, $J_{R,w}$, with the vertical scale-length of the disc reveals a positive correlation between the two, which can be approximated by the linear relation presented in Fig.~\ref{Fig_MWCorrelations}. Given that correlation does not necessarily imply causation, we interpreted that relation as consequence of disc heating, which increases the velocity dispersion raising the eccentricity of the orbits, thereby increasing the radial actions and, consequently, $J_{R,w}$. On the contrary, no clear trend is observed between the radial scale-length, $h_R$, and $J_{R,w}$, suggesting that disc heating is not the primary mechanism driving its expansion, although some influence cannot be entirely ruled out.
\begin{acknowledgements}
We thank the anonymous referee for their valuable comments, which have helped improve the revision of this manuscript. P. A. Palicio acknowledges the financial support from the Centre national d'études spatiales (CNES). 
TTG acknowledges partial financial support from the Australian Research Council (ARC) through an Australian Laureate Fellowship awarded to JBH. We have used simulations from the Auriga Project public data release \citep{Grand24} available at \url{https://wwwmpa.mpa-garching.mpg.de/auriga/data}. The NewHorizon simulation was made possible thanks to computational resources at CINES, TGCC, and KISTI, and local clusters at the Institut d'Astrophysique de Paris and the Yonsei University. This work was granted access to the HPC resources of CINES under the allocations c2016047637, A0020407637, and A0070402192 by Genci, KSC-2017-G2-0003 by KISTI, and as a `Grand Challenge' project granted by GENCI on the AMD Rome extension of the Joliot Curie supercomputer at TGCC. Although \textsc{galpy} is not explicitly used in this work, P. A. Palicio uses its source code as reference and recognises the credit for the work of \citet{Galpy}.
\end{acknowledgements}

%
%


\bibliographystyle{aa}  
\bibliography{SpiralArmsSimulations} 

\begin{appendix}
\onecolumn
\section{Maps of the potential fit}
\label{app_galpot}
\par
\begin{figure*}[!h]
\centering
\includegraphics[width=0.86\textwidth]{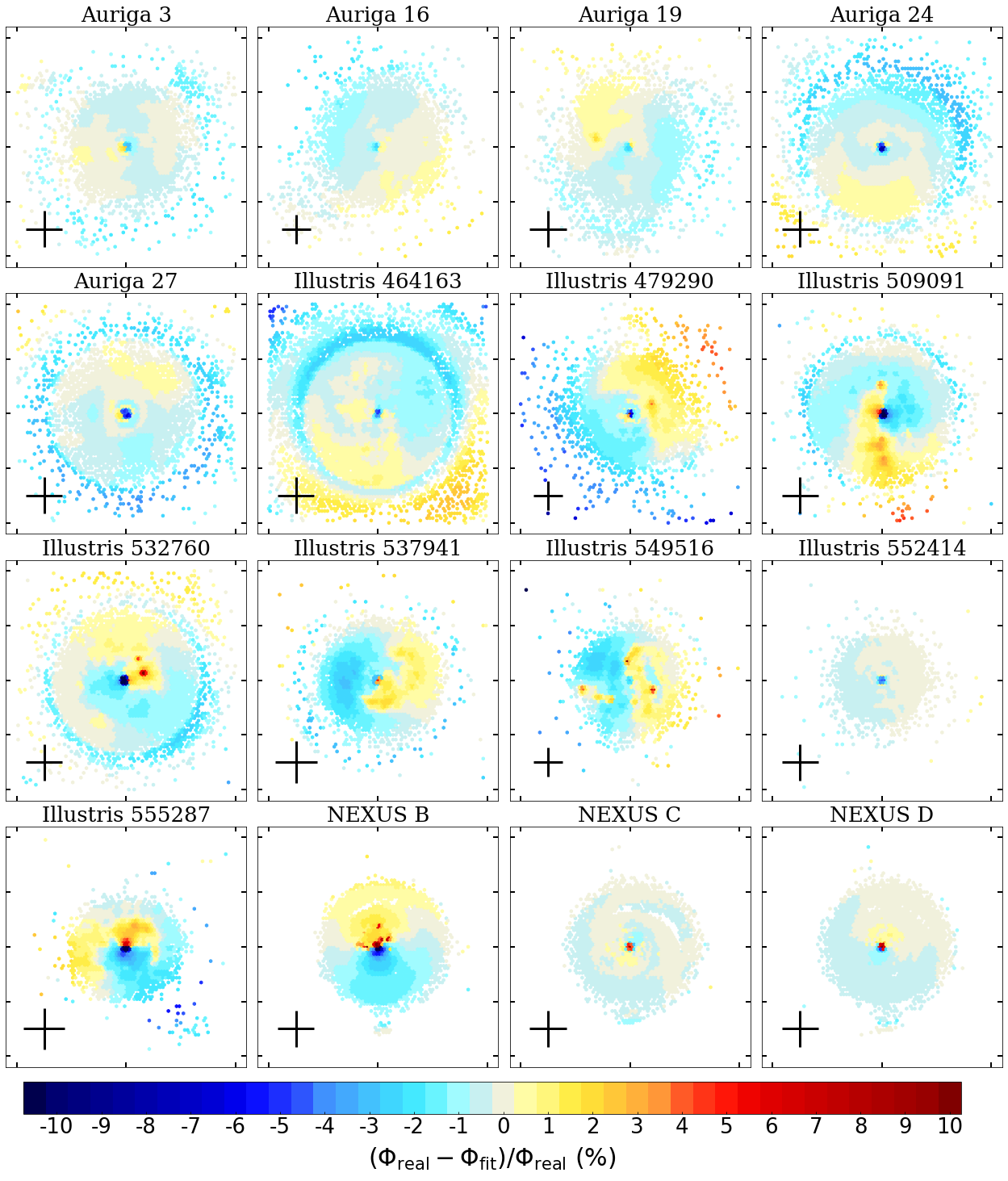}
\caption{Relative residuals of the potential fits (in percentages) for each simulation considered in this work. Each cell shows the most extreme relative discrepancy between the model and the nominal potential reported in the simulation within $|Z|<Z_{lim}$. Black cross in the lower left corner indicates the scale of $\pm 10$~kpc.}
\label{Fig_ResPot}
\end{figure*}

\begin{figure*}[phtb]
\centering
\includegraphics[width=0.97\textwidth]{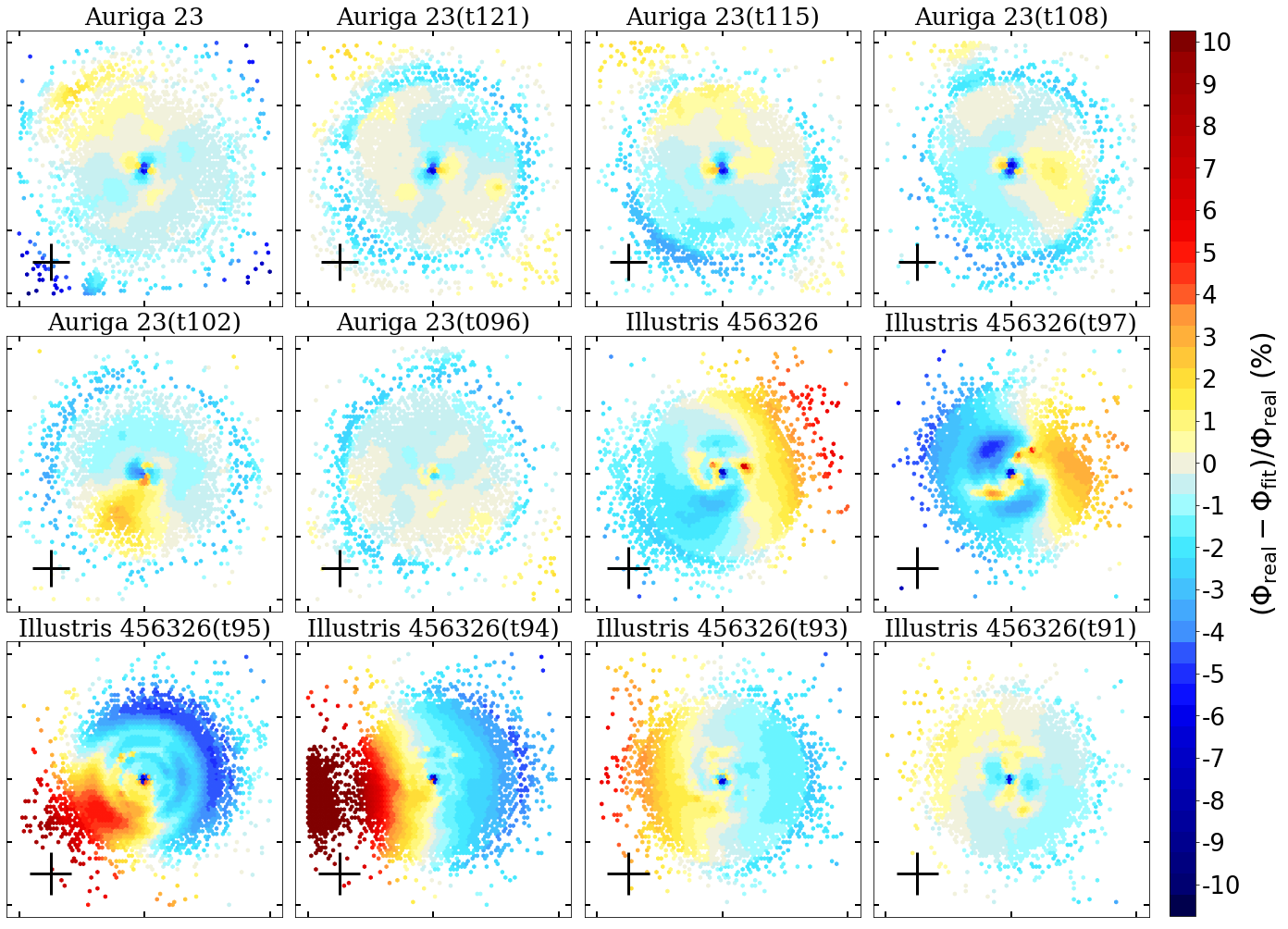}
\caption{Similar to Fig. \ref{Fig_ResPot} but for the set of \textsc{Auriga 23} and \textsc{Illustris 456326} snapshots used in Section \ref{SubSect_BadCases}. Black cross in the lower left corner indicates the scale of $\pm 10$~kpc.}
\label{Fig_ResPotMerger}
\end{figure*}

 As mentioned in Section \ref{Sect_PotFit}, for each simulation we performed a fit of the gravitational potential assuming an axisymmetric formulation for it. The most deviant relative discrepancy between this fit and the potential reported in the simulations is illustrated in Figs. \ref{Fig_ResPot} and \ref{Fig_ResPotMerger} .
\FloatBarrier
%
%
%
%
%
%
%
%
%
%
%
%
%
\section{Additional maps of density contrast and radial actions}
\par In this section, we present the maps of the density contrasts, $\delta_{\Sigma}$ and $\tilde{\delta}_{\Sigma}$, as well as those of the radial actions for the simulations not shown in Figure \ref{Fig_Actions0}.
\begin{figure*}[!p]
\centering
\includegraphics[height=0.95\textheight]{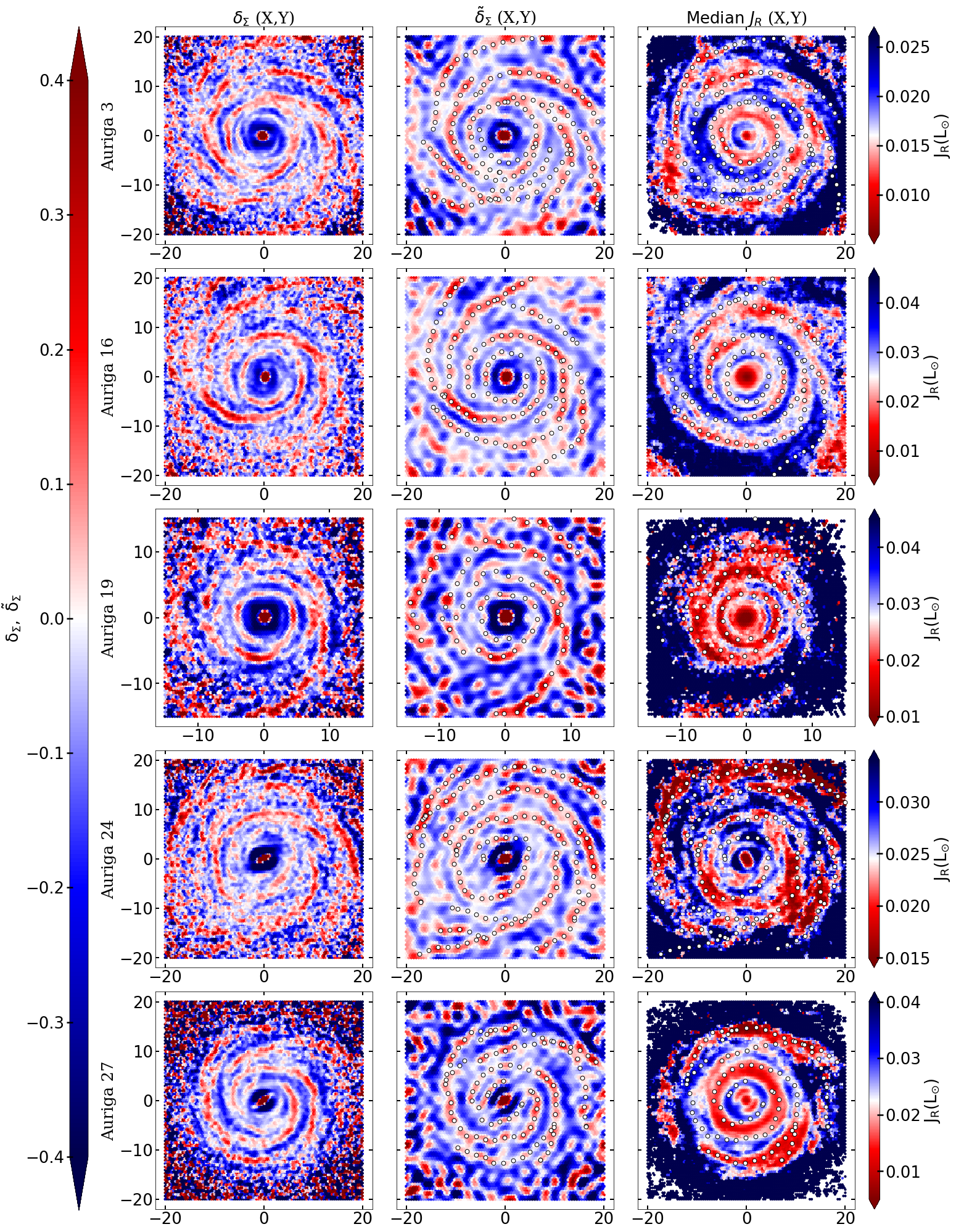}
\caption{{Maps of the density contrast (first column), its bidimensional Fourier approximation (second column), and mass-weighted median radial action (third column) for a set of simulations not shown in Fig. \ref{Fig_Actions0}.}}
\label{Fig_Actions1}
\end{figure*}
\begin{figure*}[!pthb]
\centering
\includegraphics[height=0.95\textheight]{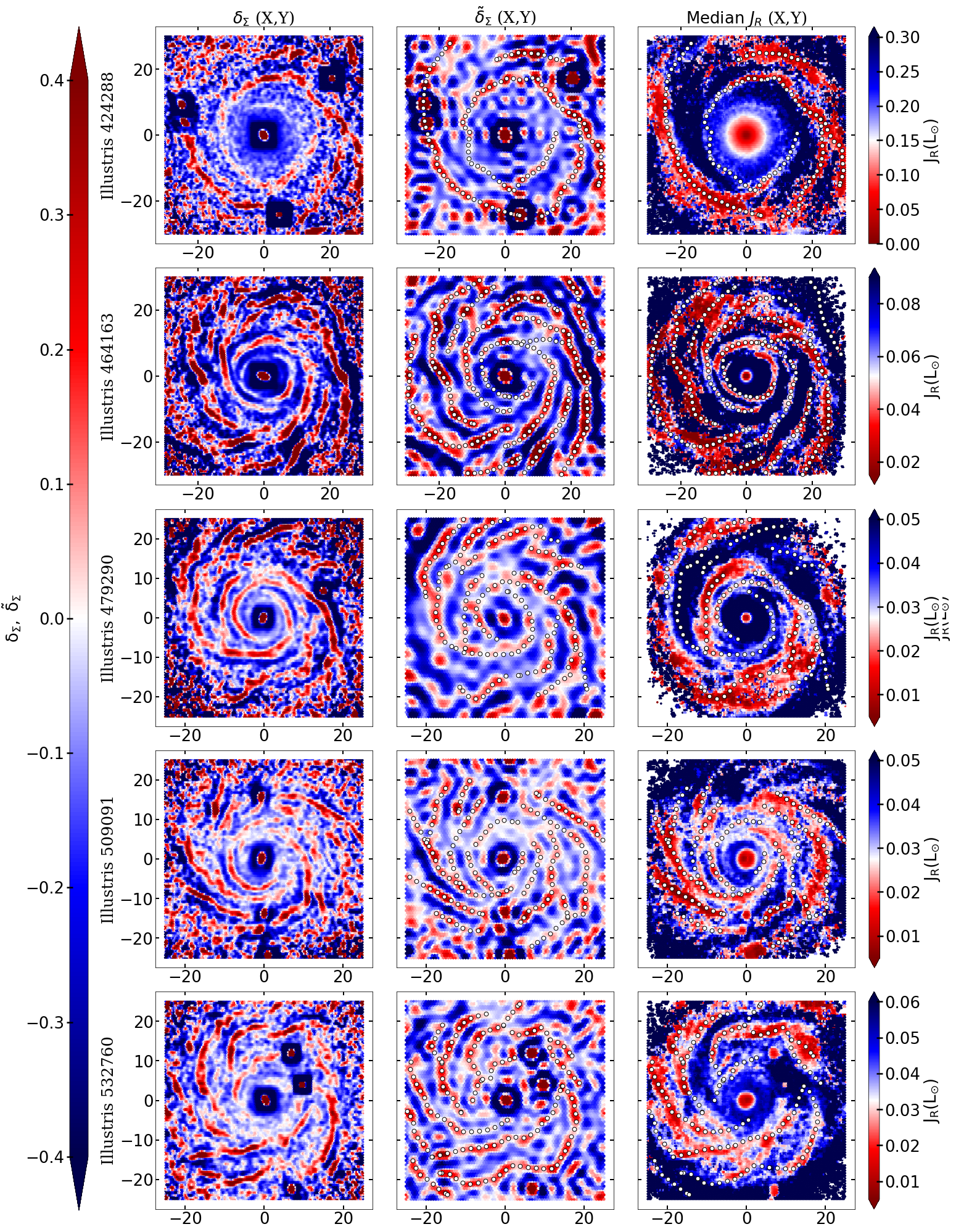}
\caption{Continuation of Fig. \ref{Fig_Actions1}.}
\label{Fig_Actions2}
\end{figure*}
\begin{figure*}[pthb]
\centering
\includegraphics[height=0.95\textheight]{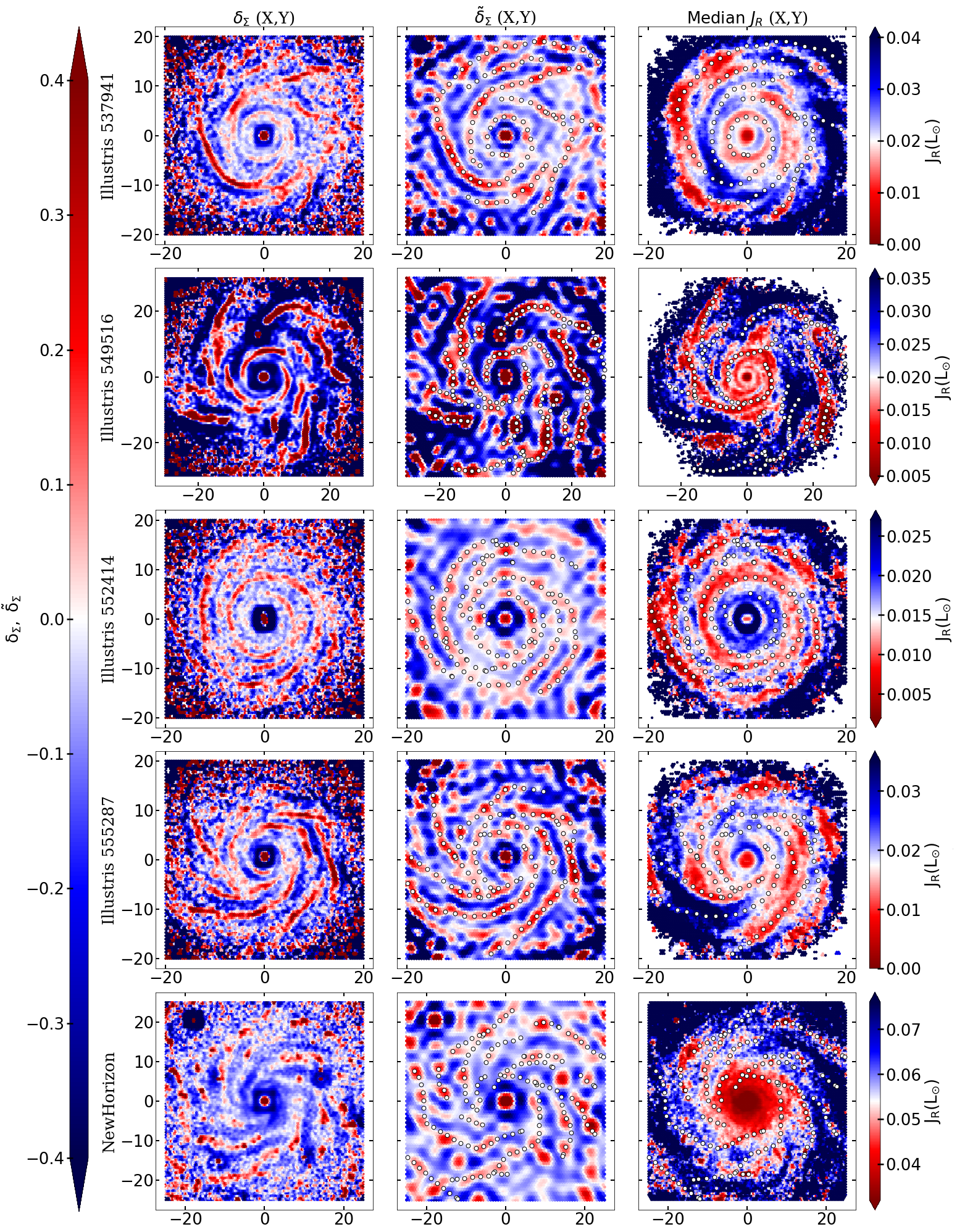}
\caption{Continuation of Fig. \ref{Fig_Actions2}.}
\label{Fig_Actions3}
\end{figure*}
\FloatBarrier
\begin{figure*}[pthb]
\centering
\includegraphics[height=0.95\textheight]{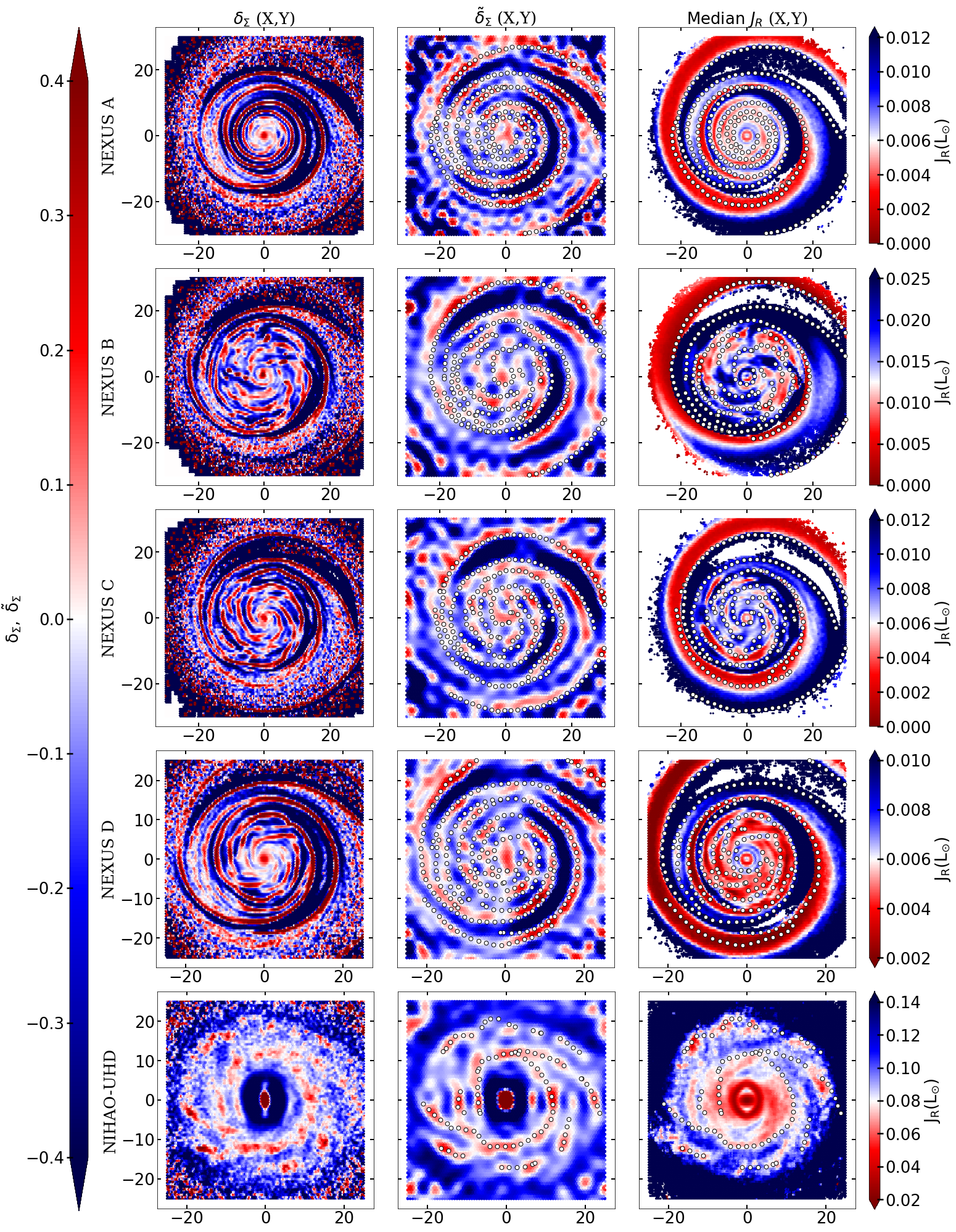}
\caption{Continuation of Fig. \ref{Fig_Actions3}.}
\label{Fig_ActionsLast}
\end{figure*}
\FloatBarrier
%
%
%
%
%
%
\section{The overdensity vs. radial action anticorrelation}
\label{app_correlation_diagram}
\par
\begin{figure*}[htp]
\centering
\includegraphics[height=0.85\textheight]{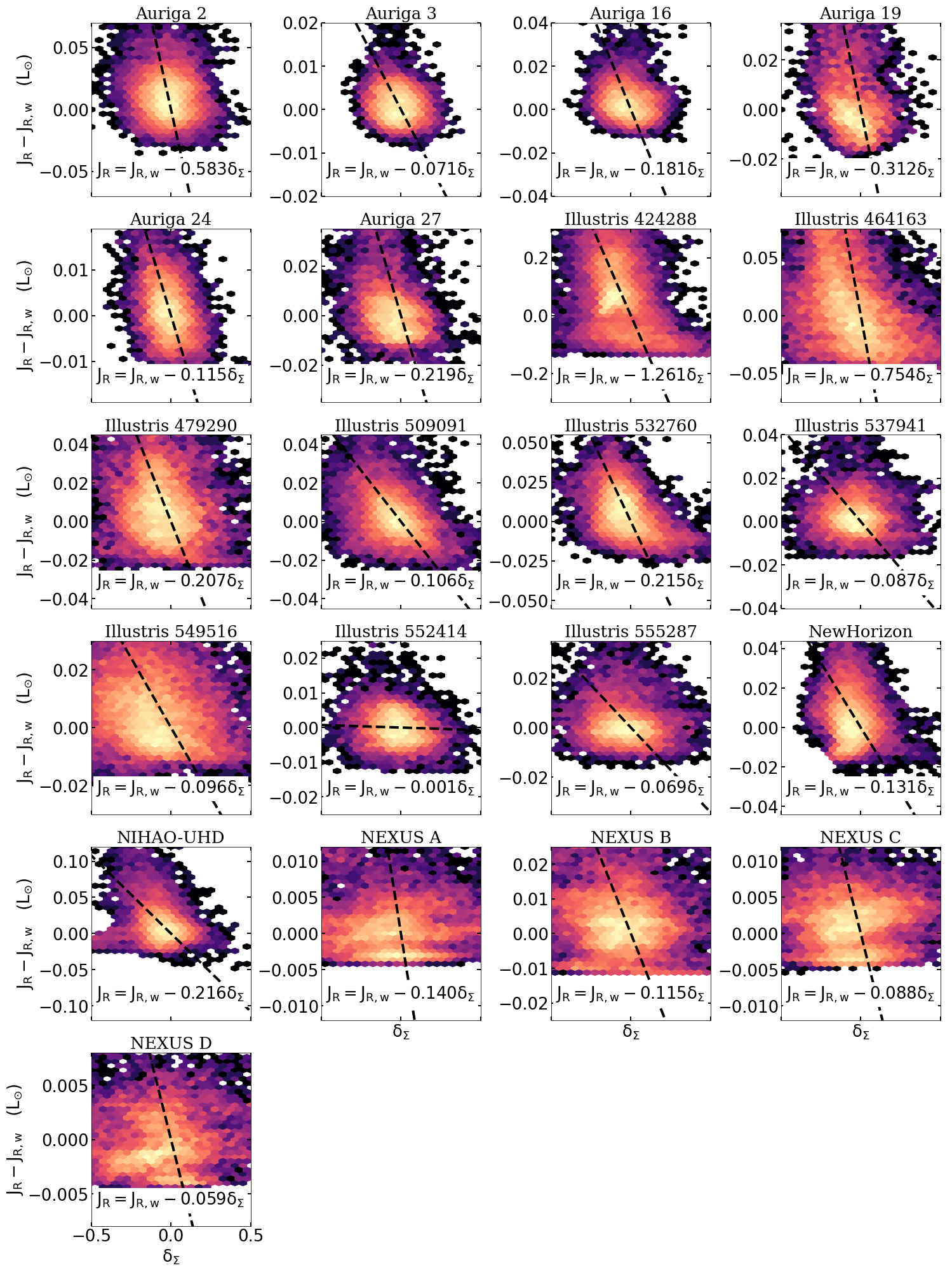}
\caption{Distribution of particles in the correlation diagram $\delta_\Sigma$ vs. $J_R$. The solid lines represent the linear fit of the particle density shown in the background map. The slope of the linear fit is provided in the inset text.}
\label{Fig_Correl0}
\end{figure*}

 {As explained in Section \ref{SubSect_Dynamics}, the difference in the width of the structures observed in the $J_R$ maps compared to those in $\delta_\Sigma(X, Y)$ hinders a clearer discernment of their relationship through visual inspection. This is evident in Figure \ref{Fig_Correl0}, where the linear fit, despite significant scatter, shows negative slopes for all the simulations.}
\FloatBarrier
%
%
%
%
%
%
%
\section{Additional age binned maps of density contrast}
\par In this section, we show the density contrast maps per bin of stellar age, $\delta_{\Sigma}(X,Y)$, for those simulations not presented in Section \ref{SubSect_Ages}.
\begin{figure*}[htbp]
\centering
\includegraphics[height=0.87\textheight]{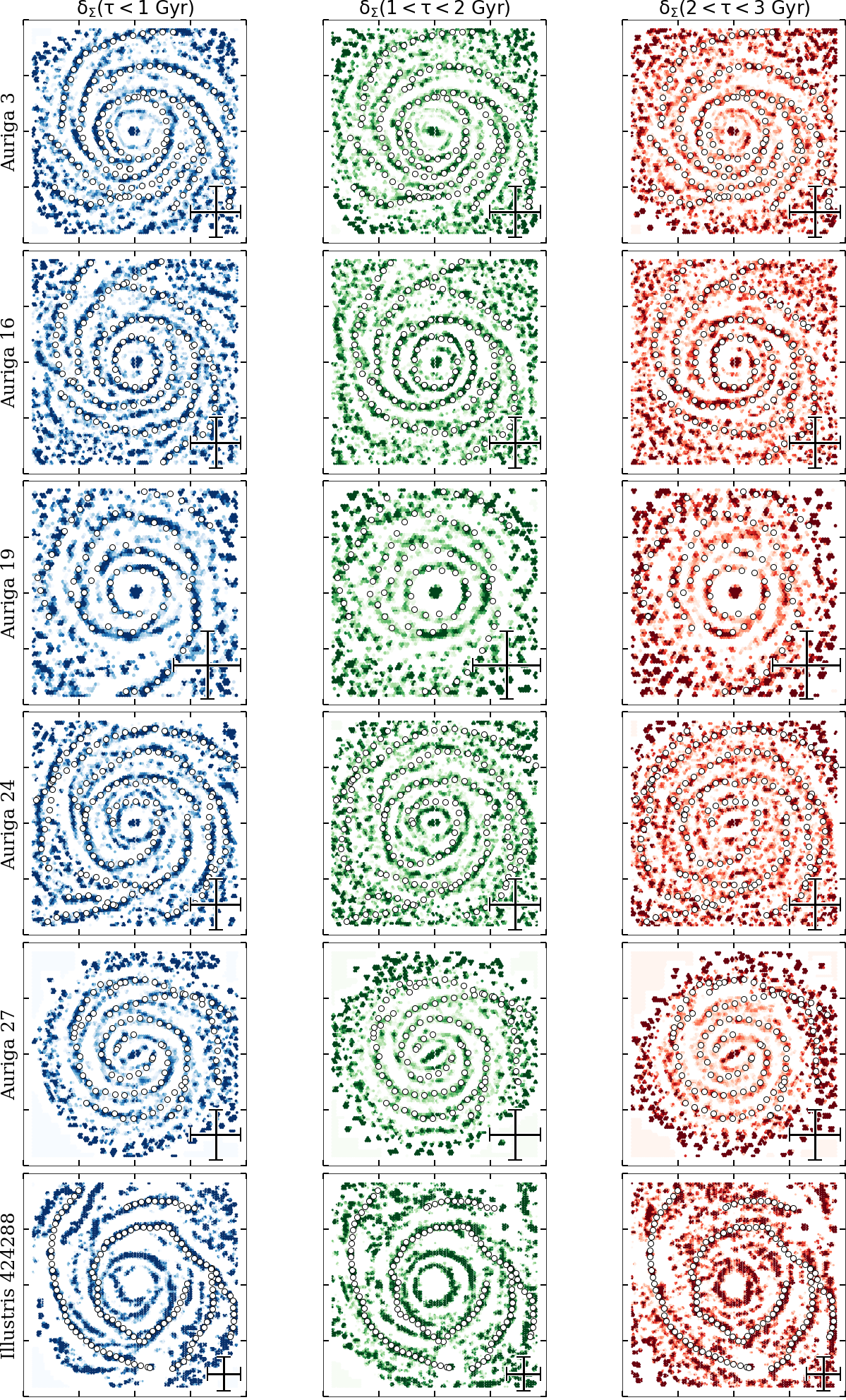}
\caption{{Density contrast maps for three distinct age intervals: stellar particles younger than 1 Gyr (left column), those aged between 1 and 2 Gyr
(middle column), and those between 2 and 3 Gyr (right column) for a set of galaxies not shown in Fig. \ref{Fig_AgeContrastCited0}. Open circles indicate the spiral arm tracks inferred from Figs. \ref{Fig_Actions1}-\ref{Fig_ActionsLast}. Cross
in the bottom-right corner indicates the scale of $\pm 5$~kpc.}}
\label{Fig_AgeContrastNonCited0}
\end{figure*}
%
\begin{figure*}[htbp]
\centering
\includegraphics[height=0.87\textheight]{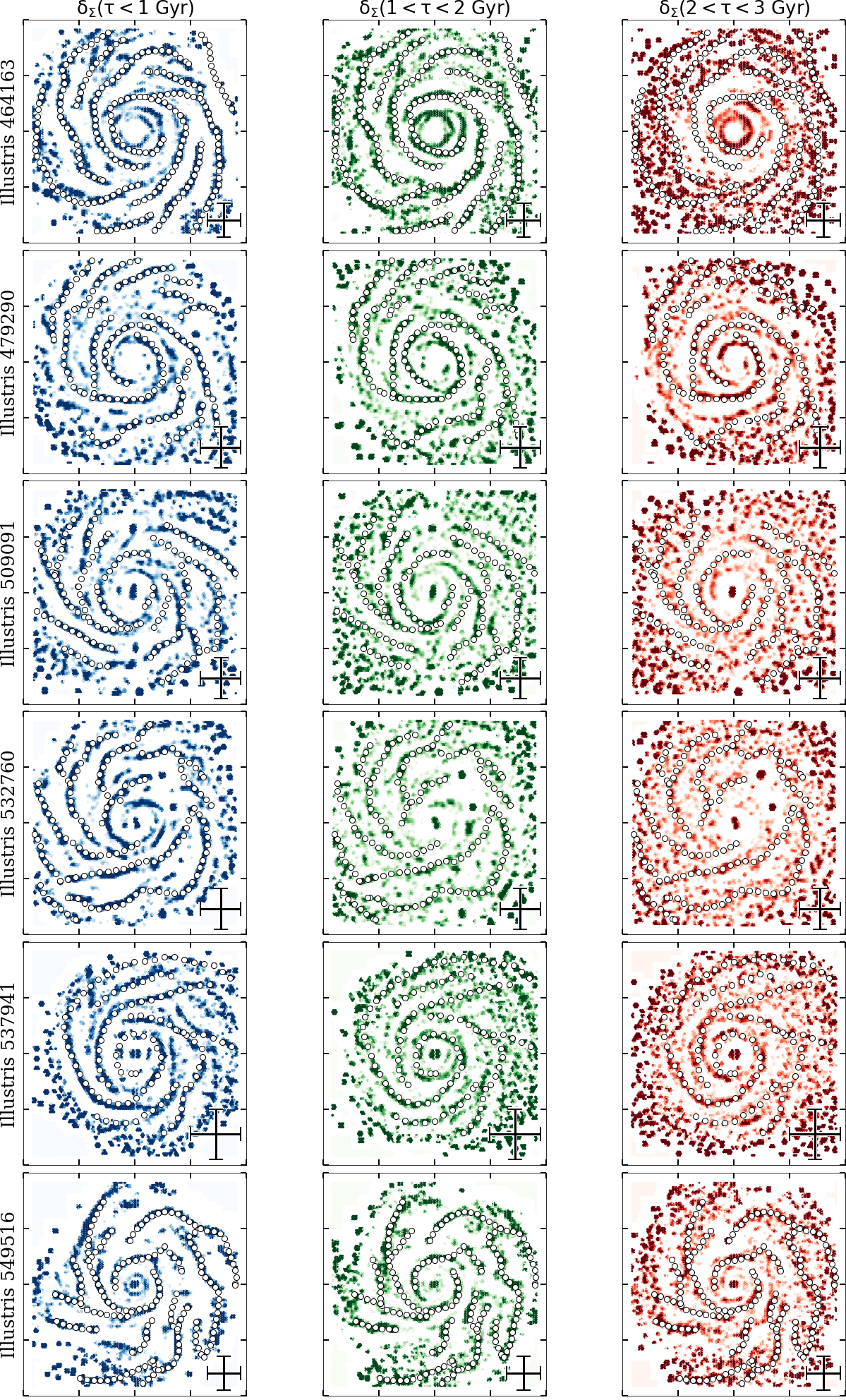}
\caption{Continuation of Fig. \ref{Fig_AgeContrastNonCited0}.}
\label{Fig_AgeContrastNonCited1}
\end{figure*}
%
\begin{figure*}[htb]
\centering
\includegraphics[height=0.57\textheight]{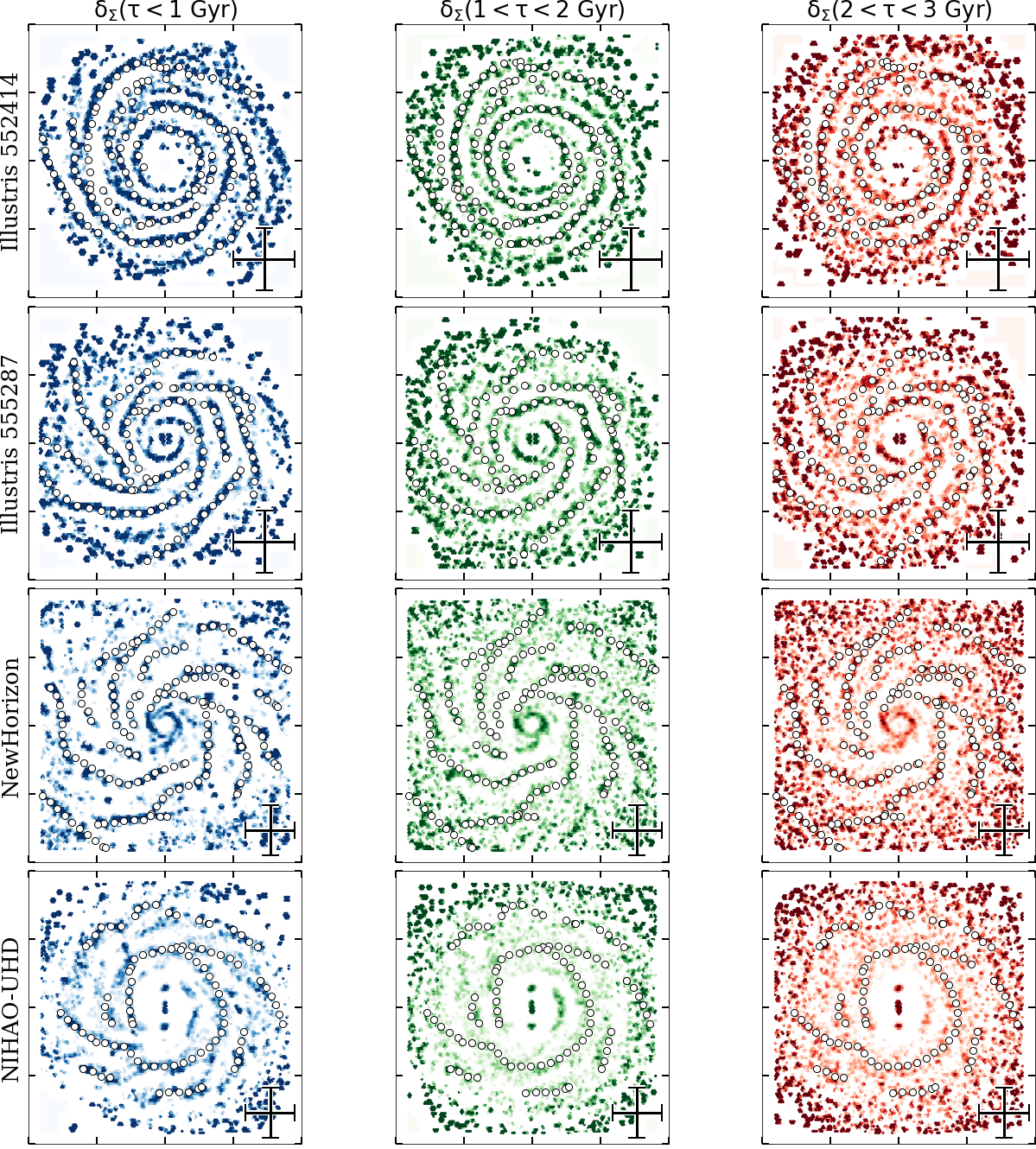}
\caption{Continuation of Fig. \ref{Fig_AgeContrastNonCited1}.}
\label{Fig_AgeContrastNonCitedLast}
\end{figure*}
\twocolumn
\section{KDE technique at discontinuities}
\label{Sec_Discontinuity}
\par
\begin{figure}[htbp]
\centering
\includegraphics[width=0.48\textwidth]{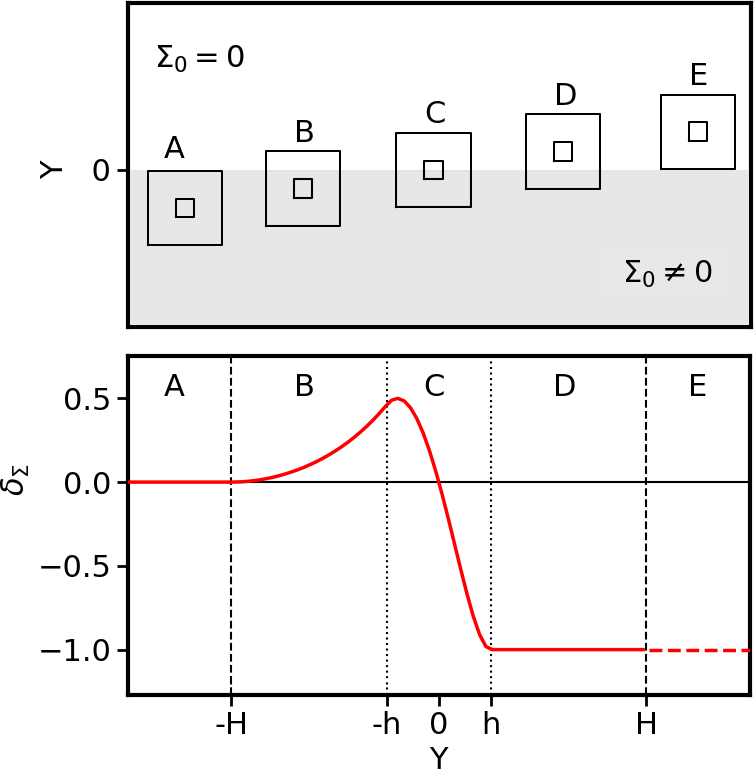}
\caption{Visualisation of the border effect produced by the KDE technique on the ${\delta}_{\Sigma}$ map of the \textsc{\ThorSimName} simulations. Upper panel: scheme of the different configurations of the local (small squares) and global (large squares) scanning windows with respect to the boundary between a medium of uniform density $\Sigma_0$ (grey area at negative Y) and the vacuum (white region at positive Y). Lower panel: values of ${\delta}_{\Sigma}$ for the mentioned configurations (red curve), assuming $h=0.5$~kpc (vertical dotted lines) and $H=2.0$~kpc (vertical dashed lines) bandwidths for the local and global kernels, respectively.}
\label{Fig_KDE_expl}
\end{figure}

 In Sect. \ref{SubSect_BadCases}, we explained a discrepancy between the ${\delta}_{\Sigma}$ and $J_R$ maps, observed in certain regions of the \textsc{\ThorSimName} simulations, as a misinterpretation of the former as a spiral arm instead of a border effect of the KDE technique. In order to illustrate the mechanisms that produces this effect, we adopted a simplified model of the scenario seen in the simulations in which a medium of constant surface density, $\Sigma_0$, limits the vacuum at $Y=0$. We can obviate the $X$ direction in this approximation. In the neighbourhood of the boundary between the two media, the estimation of ${\delta}_{\Sigma}(Y)$ depends on the distance of the kernel centre with respect to $Y=0$~kpc, as well as on the bandwidths $h$ and $H$ used for the local and global KDEs, respectively. Consequently, there are five possible configurations for the relative location of the kernel (upper panel of Fig. \ref{Fig_KDE_expl}). In the configuration labelled as A, the support of the global kernel is completely inside the dense medium. If the support of the local kernel is completely embedded in the dense medium but not that of the global kernel we obtain the configuration B. The configuration C accounts for the partial embedding of both kernel supports, while in the configuration D only that of the global kernel is partially included in the medium. Finally, in the configuration E both kernels do not overlap with the dense medium.
\par From the mathematical point of view, the generalisation of the KDE technique with continuous media implies the substitution of the summation in Eq. \ref{Eq_deltaKDE} by an integral. Therefore, the local density estimated with the Epanechnikov kernel of bandwidth $h$ is
\begin{equation}
\label{Eq_SigmaY_cont}
\Sigma(Y;h) \sim \frac{\Sigma_0}{h}\int_{c(Y,h)}^{d(Y,h)} \left(1-\frac{(y-Y)^2}{h^2}\right)dy,
\end{equation}
where the limits of the integral depend of the configurations shown in Fig. \ref{Fig_KDE_expl}. The estimation of the global density is analogous to Eq. \ref{Eq_SigmaY_cont} but replacing $h$ by $H$. The density contrast ${\delta}_{\Sigma}(Y)$ derived from these estimations is represented in the bottom panel of Fig. \ref{Fig_KDE_expl} for different distances to the interface, assuming bandwidths of $h=0.5$~kpc and $H=2.0$~kpc. As can be seen, in configuration B the estimator ${\delta}_{\Sigma}$ is positively biased while at configuration C it may result either positively  (peaking at $\sim0.5$ for this example) or negatively biased. Finally, we adopted ${\delta}_{\Sigma}=-1$ as value for the configuration E to avoid the 0/0 indeterminate form (dashed red line in Fig. \ref{Fig_KDE_expl}).
\FloatBarrier
\end{appendix}
\end{document}